\documentclass{aastex}          
\usepackage{spr-astr-addons}    

\usepackage{rotating}
\usepackage{color}

\newcommand\Tstrut{\rule{0pt}{2.6ex}}       
\newcommand\Bstrut{\rule[-0.9ex]{0pt}{0pt}} 
\newcommand{\TBstrut}{\Tstrut\Bstrut}

\begin{document}
%
\title{Optical spectroscopic classification of 35 hard X-ray sources from the Swift-BAT 70-month catalogue}

\shorttitle{Optical classification of 35 Swift hard X-ray sources}
\shortauthors{E. J. Marchesini et al.}

\author{E. J. Marchesini \altaffilmark{1,2,3,4,5}}
\email{ejmarchesini@gmail.com}
   \and 
\author{N. Masetti \altaffilmark{5,6}}
   \and 
\author{E. Palazzi\altaffilmark{5}}
   \and 
\author{V. Chavushyan\altaffilmark{7}}
   \and 
\author{E. Jim\'enez-Bail\'on\altaffilmark{8}}
   \and 
\author{V. M. Pati\~{n}o-\'Alvarez\altaffilmark{9,7}}
   \and 
\author{V. Reynaldi\altaffilmark{1,2}}
   \and 
\author{A. F. Rojas\altaffilmark{6,10}}
   \and 
\author{I. Saviane\altaffilmark{10}}
   \and 
\author{I. Andruchow\altaffilmark{1,2}}
   \and 
\author{L. Bassani\altaffilmark{5}}
   \and 
\author{A. Bazzano\altaffilmark{11}}
   \and 
\author{A. J. Bird\altaffilmark{12}}
   \and 
\author{A. Malizia\altaffilmark{5}}
   \and 
\author{D. Minniti\altaffilmark{6,13,14}}
   \and 
\author{L. Monaco\altaffilmark{6}}
   \and 
\author{J. B. Stephen\altaffilmark{5}}
   \and 
\author{P. Ubertini\altaffilmark{11}}


\altaffiltext{1}{Facultad de Ciencias Astron\'omicas y Geof\'isicas, Universidad Nacional de La Plata, Paseo del Bosque, B1900FWA, La Plata, Argentina}
\altaffiltext{2}{Instituto de Astrof\'isica de La Plata, CONICET--UNLP, CCT La Plata, Paseo del Bosque, B1900FWA, La Plata, Argentina}
\altaffiltext{3}{Dipartimento di Fisica, Universit\`a degli Studi di Torino, via Pietro Giuria 1, I-10125 Turin, Italy}
\altaffiltext{4}{INFN -- Istituto Nazionale di Fisica Nucleare, Sezione di Torino, via Pietro Giuria 1, I-10125 Turin, Italy}
\altaffiltext{5}{INAF -- Osservatorio di Astrofisica e Scienza dello Spazio, via Gobetti 93/3, I-40129, Bologna, Italy}
\altaffiltext{6}{Departamento de Ciencias Fisicas, Universidad Andres Bello, Fernandez Concha 700, Las Condes, Santiago, Chile}
\altaffiltext{7}{Instituto Nacional de Astrof\'isica, \'Optica y Electr\'onica, Apartado Postal 51-216, 72000 Puebla, Mexico}
\altaffiltext{8}{Universidad Nacional Auto\'onoma de M\'exico, Apartado Postal 70-264, 04510 M\'exico D.F., Mexico}
\altaffiltext{9}{Max-Planck-Institut f\"ur Radioastronomie, Auf dem H\"ugel 69, 53121 Bonn, Germany}
\altaffiltext{10}{European Southern Observatory, Alonso de Cordova 3107, Vitacura, Santiago, Chile}
\altaffiltext{11}{INAF--Istituto di Astrofisica e Planetologia Spaziali, Via Fosso del Cavaliere 100, I-00133 Rome, Italy}
\altaffiltext{12}{Physics \& Astronomy, University of Southampton, Southampton, Hampshire, SO17 1BJ, United Kingdom}
\altaffiltext{13}{Millennium Institute of Astrophysics, Av. Vicu\~{n}a Mackenna 460, Macul, Santiago, Chile}
\altaffiltext{14}{Vatican Observatory, 00120 Vatican City State}

\begin{abstract}
The nature of a substantial percentage (about one fifth) of hard X-ray sources discovered with the BAT instrument onboard the \emph{Neil Gehrels Swift Observatory} (hereafter Swift) is unknown because of the lack of an identified longer-wavelength counterpart. Without such follow-up, an X-ray catalogue is of limited astrophysical value: we therefore embarked, since 2009, on a long-term project to uncover the optical properties of sources identified by Swift by using a large suite of ground-based telescopes and instruments.

In this work, we continue our programme of characterization of unidentified or poorly studied hard X-ray sources by presenting the results of an optical spectroscopic campaign aimed at pinpointing and classifying the optical counterparts of 35 hard X-ray sources taken from the 70-month BAT catalogue. This sample was selected out of the available information about the chosen objects: either they are completely unidentified sources, or their association with a longer-wavelength counterpart is still ambiguous.

With the use of optical spectra taken at six different telescopes we were able to identify the main spectral characteristics (continuum type, redshift, and emission or absorption lines) of the observed objects, and determined their nature.

We identify and characterize a total of 41 optical candidate counterparts corresponding to 35 hard X-ray sources given that, because of positional uncertainties, multiple lower energy counterparts can sometimes be associated with higher energy detections. We discuss which ones are the actual (or at least most likely) counterparts based on our observational results.

In particular, 31 sources in our sample are active galactic nuclei: 16 are classified as Type 1 (with broad and narrow emission lines) and 13 are classified as Type 2 (with narrow emission lines only); two more are BL Lac-type objects. We also identify one LINER, one starburst, and 3 elliptical galaxies. The remaining 5 objects are galactic sources: we identify 4 of them as cataclysmic variables, whereas one is a low mass X-ray binary.
\end{abstract}


\maketitle
   
%

\section{Introduction}

The number of hard X-ray and $\gamma$-ray detected sources has been continuously increasing since the advent of large field of view space telescopes capable of observing in such energy bands, and the subsequent increment of sensitivity and positional accuracy with each new technological generation. In particular, the instruments IBIS \citep{Ubertini03} and BAT \citep{Barthelmy05}, onboard the INTEGRAL \citep{Winkler03} and Swift \citep{Gehrels04} satellites respectively, provide hard X-ray detections with a positional uncertainty of less than 5 arcminutes. Today there are more than 1500 hard X-ray sources listed as detected by these satellites \citep[e.g.,][]{Bird16,Oh18}. Of these, around 60\% are classified as active galactic nuclei (AGN), 15\% as X-ray binaries (XRBs) and 5\% as cataclysmic variables (CVs). The remaining $\sim$20\% has not been classified yet, mainly due to one of the two following reasons: the lack of spectral information for their optical/near-infrared counterpart, or the lack of a clear lower energy counterpart to associate them with. However it is mainly through the characterization of its spectral shape and features, in particular in the optical, that we can confirm any kind of associated counterpart, and study the nature of these unclassified objects.

The study and classification of unidentified high energy sources is of key importance to a number of relevant fields: population analysis of extragalactic objects \citep{Ricci17}, binary system studies \citep{Coleiro13}, the hard X-ray emitting AGN luminosity function \citep{Sazonov07}, systematic studies of nearby absorbed AGN \citep[][and references therein]{Malizia12}, among others. Besides, it has been proven that X-ray observations are useful for finding locally-obscured AGN \citep{Koss16a}, new types of X-ray binaries, as the supergiant fast x-ray transients \citep[SFXTs,][]{Sguera06}, and $\gamma$-ray binaries \citep{deMartino13,Strader16}. 

Most of the unidentified X-ray sources as seen by INTEGRAL/IBIS were identified by our group \citep[][and references therein]{Masetti13}. In the same fashion, there had been a number of works regarding the analysis and classification of BAT unidentified sources via optical spectroscopy \citep[e.g.,][]{Landi07,Parisi09,Berney15,Halpern15,Koss17}, but the classification task remained incomplete, mainly due to the fact that the fainter optical counterparts are harder to observe, especially those in the southern hemisphere where a smaller number of medium-sized (4-metre) telescopes is available. Here we continue the work started in \citet{Parisi09} and continued in \citet{Parisi12,Parisi14} and in \citet{Rojas17}, focusing in particular on the identification and classification of unidentified hard X-ray sources as listed in the 70-months BAT All-sky Hard X-ray Survey catalogue \citep[][hereafter Swift70]{Baumgartner13}. We then compare our results with the newest version of the survey, the BAT 105-months All-sky Hard X-ray Survey catalogue \citep[][hereafter Swift105]{Oh18}.

The paper is structured as follows: in Sect. 2, we present and discuss the sample selection; in Sect. 3 we describe the observations, telescopes, instruments and reduction techniques used; in Sect. 4 we present the classification of the observed sources and their main characteristics; in Sect. 5 we separately discuss cases of interest within the sample (namely, those with multiple longer-wavelength counterparts for a single BAT detection); and in Sect. 6 we draw the main conclusions of our work.

\section{Sample selection}

The Swift70 catalogue displays a full list of observational characteristics for each detected object. In particular, it shows a 'Type' designator for each one, when the source is unambiguously classified through a cross-match with the Simbad database \citep{Wenger00} as a means of providing the proposed counterpart. However, as stated in the previous Section, in about 20\% of the cases either no association with other objects is found, or the associated counterpart is an unidentified/unclassified object: these are, in all, hard X-ray sources with no clear classification. 

Our subsample was thus selected from this catalogue on the basis of the following criteria:

\begin{enumerate}
\item detected hard X-ray sources with no associated counterpart, or
\item detected hard X-ray sources with an associated counterpart which has not been classified up to now.
\end{enumerate}

In the cases for which there was no known counterpart in the catalogue, we first performed a search in the soft X-ray data archives available online, in particular in the Swift/XRT 1SXPS \citep{Evans14} and XMM-Newton XMMSSC \citep{Rosen16} catalogs. This search was centered at the BAT equatorial coordinates and done within a circular area with a radius equal to the BAT positional uncertainty for each given source. We selected every soft X-ray source detected within the BAT positional uncertainty and with appreciable emission above 3 keV (i.e. with a signal-to-noise ratio greater than 3) as a candidate counterpart.

This procedure, according to \citet{Stephen06}, allows the association of longer-wavelength counterparts to high-energy sources with high degree of confidence and, therefore, reduces the positional uncertainty from the usual one for BAT (a few arcminutes) to that typical of soft X-ray data (a few arcseconds), thus shrinking the search area by a factor of up to 10$^4$. Then, using the positions and (smaller) uncertainties obtained from the soft X-ray data, we searched for optical counterparts within this smaller error circle in the DSS\footnote{Based on photographic data obtained using The UK Schmidt Telescope. The UK Schmidt Telescope was operated by the Royal Observatory Edinburgh, with funding from the UK Science and Engineering Research Council, until 1988 June, and thereafter by the Anglo-Australian Observatory. Original plate material is copyright (c) of the Royal Observatory Edinburgh and the Anglo-Australian Observatory. The plates were processed into the present compressed digital form with their permission. The Digitized Sky Survey was produced at the Space Telescope Science Institute under US Government grant NAG W-2166 and is available at {\tt http://archive.eso.org/dss/dss .}} archive. It is worth noting that there is not always a one-to-one relation between hard X-ray sources and soft X-ray counterparts, or between soft X-ray sources and optical counterparts. In some cases, more than one soft X-ray source lies within the hard X-ray positional uncertainty circle, and the same can happen with multiple optical sources lying within a soft X-ray positional uncertainty circle: this is why the number of optical sources studied here and proposed as counterparts is larger than the total number of hard X-ray sources under scrutiny in this work (see Tables 1 and 2, and Sect. 4). With this approach, for each BAT source, we selected one or few (three at most) putative optical counterparts with magnitude $R \leq$ 21 to be spectroscopically observed. In the end, we obtained a spectrum for every optical counterpart (8 in total, corresponding to 7 hard X-ray sources) pinpointed in this way.

Furthermore, after having checked the positional correspondence with a catalogued soft X-ray source, we also obtained spectra for 33 optical sources which were already proposed as counterparts to 28 hard X-ray sources in the Swift70 catalogue, albeit ill-studied or of unidentified nature. These counterparts, according to the latter survey, have been selected following the recipe detailed in \citet{Tueller10}. In summary, and in a similar fashion to our method, they crossmatched the BAT positions with soft X-ray data archives. If there was no soft X-ray data available, then a proposal was made to observe the field with XRT for 10 kiloseconds to find a soft X-ray counterpart. The better positional accuracy ($\sim$6 arcseconds) of these soft X-ray counterparts was then used to associate one optical counterpart to each, by crossmatching them with the NED and SIMBAD databases. If, however, no soft X-ray counterpart was found this way, then a counterpart candidate selection was made based only on the nature of sources available in the NED and SIMBAD databases within the corresponding field. This procedure was then classified with 8 different association strength values, from 0 to 7, being 0 the strongest association, 7 the weakest \citep[see table 5 in][]{Baumgartner13}. In the case of optical crowded fields, more than one counterpart could be selected.

We stress that all BAT objects of our sample could be positionally associated with at least one soft X-ray counterpart with the exception of Swift J0550.7-3215 B, for which we could not find a corresponding soft X-ray source within the hard X-ray error circle. We nevertheless decided to spectroscopically study the corresponding candidate optical counterpart proposed in Swift70. In total, we selected a subsample of 35 sources detected by BAT, for which there are 41 optical candidate counterparts. In Table 1, we present our sample of 35 Swift/BAT sources: In column 1 we report the name as listed in the Swift70 catalog, in columns 2 and 3 the equatorial celestial coordinates referred to the J2000.0 epoch, in column 4 the BAT positional uncertainty radius, in column 5 the signal to noise ratio of the corresponding BAT detection, and in column 6 the flag indicating the strength of the counterpart association, coded according to Swift70.

Within our sample of 35 BAT detections, 23 of them show the presence of a single soft X-ray source within their hard X-ray error circle, which in turn is associated with a single optical counterpart.

On the other hand, 12 BAT sources are associated with multiple soft X-ray and/or optical counterparts: 2 sources in 11 cases, and 3 objects for the remaining one (Swift J0107.7-1137), for a total of 25 instances. In all these cases, Swift70 provides multiple optical counterparts for a single BAT source, while Swift105 provides one resolved BAT detection for each optical counterpart association. Moreover, out of these 12 BAT detections, 3 are not only associated with multiple optical sources, but they are also interacting with each other: the optical images available in the literature or in on-line databases reveal typical galaxy merging structures. Out of the 25 optical sources proposed as counterparts for these 12 BAT detections, 7 were already identified in Swift70 (therefore they were not considered in the present work): we here provide results for the remaining 18. The differences between the Swift70 and Swift105 catalogs may reside in the longer total exposure of the latter one, and in the low signal to noise ratio of the unidentified sources, since they report that the data reduction and source detection procedures implemented in both catalogs are the same.

For the sake of completeness, we also included in our sample the source PBC J0819.2-2508, for which \citet{Rojas17} could obtain the optical spectrum of only one of the two putative optical candidates associated with the hard X-ray emission detected by \citet{Cusumano10}. Here we thus report on the optical candidate associated with the soft X-ray source within the Swift/BAT error circle of PBC J0819.2-2508 showing emission above 3 keV.

\begin{table*}
\caption{The selected sample of 35 sources from the Swift70 catalogue. In column 1 we report the source name as listed in the catalogue, in columns 2 and 3 their equatorial celestial coordinates, in column 4 their associated positional uncertainty, in column 5 the signal to noise ratio of the corresponding detection and in column 6 the association flag as reported in Swift70. The latter is such that a 0 corresponds to the strongest association, and a 7 to no association available. In the case of the only source taken from the 2PBC catalog, we considered the brighter X-ray source reported in \citet{Rojas17} (see text).}
\small
\centering
\begin{tabular}{|l|c|c|c|c|c|}
\hline\hline
  \multicolumn{1}{|c|}{Object} &
  \multicolumn{1}{c|}{RA (Swift)} &
  \multicolumn{1}{c|}{DEC (Swift)} &
  \multicolumn{1}{c|}{BAT err.} &
  \multicolumn{1}{c|}{BAT} &
  \multicolumn{1}{c|}{BAT} \\
  \multicolumn{1}{|c|}{} &
  \multicolumn{1}{c|}{(J2000.0)} &
  \multicolumn{1}{c|}{(J2000.0)} &
  \multicolumn{1}{c|}{[arcmin]} &
    \multicolumn{1}{c|}{S/N} &
  \multicolumn{1}{c|}{Ctpt. flag} \\
\hline\hline
  Swift J0042.9+3016 A  & 00:42:59.04 & +30:15:32.4 & 3.93 & 7.1 & 0\\
  Swift J0052.3-2730    & 00:52:10.56 & -27:28:01.2 & 5.57 & 5.0 & 0\\
 Swift J0107.7-1137 A/B/C & 01:07:46.32& -11:37:19.2 & 4.1 & 6.8 & 0\\
  Swift J0136.5+3906    & 01:36:29.52 & +39:09:07.2 & 5.78 & 4.9 & 0\\
  Swift J0138.8+2925    & 01:38:42.48 & +29:26:24.0 & 5.72 & 32.9 & 0\\
  Swift J0222.3+2509    & 02:22:22.32 & +25:09:18.0 & 5.77 & 4.9 & 0\\
  Swift J0428.2-6704 A  & 04:28:16.32 & -67:05:27.6 & 5.79 & 4.8 & 0\\
  Swift J0449.6-5515    & 04:49:47.04 & -55:14:52.8 & 5.75 & 4.9 & 0\\
  Swift J0503.7-2819    & 05:03:45.60 & -28:20:31.2 & 5.61 & 5.0 & 0\\
  Swift J0528.1-3933    & 05:28:05.52 & -39:33:10.8 & 4.80 & 5.8 & 4\\
  Swift J0535.2+2830    & 05:35:06.48 & +28:29:06.0 & 5.64 & 5.0 & 0\\
  Swift J0550.7-3215 B  & 05:50:40.56 & -32:15:00.0 & 1.62 & 17.6 & 0\\
  Swift J0609.5-6245 B  & 06:09:46.80 & -62:42:00.0 & 4.87 & 5.8 & 0\\
  Swift J0659.3+2406    & 06:59:20.64 & +24:06:10.8 & 5.51 & 5.1 & 0\\
  Swift J0701.3-3235    & 07:04:23.28 & +26:25:15.6 & 5.46 & 4.9 & 0\\
  Swift J0725.8+3000    & 07:25:48.00 & +29:59:56.4 & 5.80 & 4.8 & 0\\
  Swift J0733.9+5156    & 07:34:05.52 & +51:54:07.2 & 5.69 & 4.9 & 0\\
  Swift J0755.4+8402    & 07:56:11.76 & +83:58:08.4 & 5.46 & 5.1 & 0\\
  Swift J1007.4+6534    & 10:06:45.12 & +65:33:43.2 & 5.50 & 5.1 & 0\\
  Swift J1041.4-1740    & 10:41:32.40 & -17:37:12.0 & 4.62 & 6.1 & 0\\
  Swift J1051.2-1704 B  & 10:51:14.40 & -17:03:03.6 & 3.59 & 7.8 & 0\\
  Swift J1114.3+2020    & 11:14:07.20 & +20:22:19.2 & 5.61 & 5.0 & 6\\
  Swift J1306.4-4025 B  & 13:06:39.12 & -40:24:14.4 & 1.67 & 17.1 & 0\\
  Swift J1512.2-1053 B  & 15:12:07.44 & -10:49:37.2 & 5.23 & 5.4 & 0\\
  Swift J1643.2+7036    & 16:42:53.04 & +70:38:34.8 & 5.55 & 5.1 & 0\\
  Swift J1731.3+1442    & 17:31:13.44 & +14:43:58.8 & 5.77 & 4.9 & 0\\
  Swift J1825.7+7215    & 18:25:43.92 & +72:12:21.6 & 5.65 & 5.0 & 0\\
  Swift J1852.8+3002    & 18:52:47.28 & +30:03:14.4 & 5.10 & 5.5 & 0\\
  Swift J1925.0+5041    & 19:25:00.00 & +50:41:09.6 & 5.53 & 5.1 & 0\\
  Swift J2024.0-0246    & 20:24:01.44 & -02:45:18.0 & 5.66 & 5.0 & 0\\
Swift J2059.6+4301 A/B  & 20:59:31.44 & +42:59:52.8 & 5.84 & 4.8 & 0\\
  Swift J2237.0+2543    & 22:36:34.08 & +25:45:03.6 & 5.53 & 5.1 & 0\\
  Swift J2242.4-3711    & 22:42:26.88 & -37:07:01.2 & 5.78 & 4.9 & 0\\
  Swift J2352.6-1707    & 23:52:47.76 & -17:05:31.2 & 5.41 & 5.2 & 0\\
\hline
  2PBC J0819.2-2508    & 08:19:15.84 & -25:08:49.2 & 4.99 & 5.0 & -\\
\hline\hline\end{tabular}
\end{table*}

\begin{table*}
\caption{The selected sample of 41 optical candidate counterparts for our sample of 35 Swift70 sources. Unless stated otherwise, the coordinates in columns 2 and 3 are extracted from the 2MASS catalogue. Optical coordinates marked with $\dagger$ are taken from the USNO-A2.0 catalogue and coordinates marked with $\ddagger$ are taken from the USNO-B1.0 catalogue. Objects whose redshift was not available in the literature until our measurements are marked with $\star$.}
\small
\centering
\setlength{\tabcolsep}{1.5pt}
\begin{tabular}{|l|c|c|c|c|c|c|c|}
\hline\hline
  \multicolumn{1}{|c|}{Object} &
  \multicolumn{1}{c|}{RA (Ctpt.)} &
  \multicolumn{1}{c|}{DEC (Ctpt.)} &
  \multicolumn{1}{c|}{Tel.+Inst.} &
  \multicolumn{1}{c|}{UT date} &
  \multicolumn{1}{c|}{Exp. time} &
  \multicolumn{1}{c|}{$\lambda$ range} &
  \multicolumn{1}{c|}{Res.} \Tstrut \\
    \multicolumn{1}{|c|}{} &
    \multicolumn{1}{c|}{(J2000.0)} &
  \multicolumn{1}{c|}{(J2000.0)} &
  \multicolumn{1}{c|}{} &
  \multicolumn{1}{c|}{[YYYY-MM-DD]} &
  \multicolumn{1}{c|}{N$\times$[s]} &
  \multicolumn{1}{c|}{[\AA]} &
  \multicolumn{1}{c|}{[\AA]} \Bstrut \\

\hline
  Swift J0042.9+3016 A$^{\star}$  & 00:43:01.84  &  30:17:19.6 & Cassini-BFOSC & 2013-01-07 & 2$\times$1800 & 3400-8550 & 13.5 \\
  Swift J0052.3-2730$^{\star}$    & 00:52:03.83 & -27:23:48.9 & TNG-DOLoReS & 2013-09-19 & 2$\times$1200 & 3700-3900 & 14.5\\
  Swift J0107.7-1137 A$^{\star}$  & 01:07:37.83$^\dagger$  &  -11:39:57.1$^\dagger$ & TNG-DOLoReS & 2012-08-20 & 2$\times$1800 & 3700-3900 & 14.5\\
  Swift J0107.7-1137 B            & 01:07:39.64  &  -11:39:11.8 & UKST-SuperCOSMOS I & 2004-09-12 & 1200+600 & 3900-7600 & 21.5\\
  Swift J0107.7-1137 C$^{\star}$  & 01:07:13.77  &  -11:36:02.8 & SPM-B\&C & 2012-12-03 & 3$\times$1800 & 3300-7900 & 11.2\\
  Swift J0136.5+3906              & 01:36:32.61$^\dagger$  &  39:05:59.2$^\dagger$ & Cassini-BFOSC & 2013-01-07 & 2$\times$1800 & 3400-8550 & 12.7\\
  Swift J0138.8+2925              & 01:39:24.00  & 29:24:06.8 & Cassini-BFOSC & 2013-09-23 & 2$\times$1800 & 3400-8550 & 13.5\\
  Swift J0222.3+2509              & 02:22:35.23  &  25:08:14.3 & Cassini-BFOSC & 2013-01-07 & 2$\times$1200 & 3400-8550 & 12.7\\
  Swift J0428.2-6704 A            & 04:27:49.60$^\dagger$  &  -67:04:34.9$^\dagger$  & NTT-EFOSC & 2017-02-17 & 2$\times$900 & 4000-8500 & 18.1\\
  Swift J0449.6-5515$^{\star}$    & 04:50:01.94  &  -55:12:40.5 & NTT-EFOSC & 2017-02-17 & 3$\times$10 & 4000-8500 & 15.5\\
  Swift J0503.7-2819$^{\star}$    & 05:03:33.07  & -28:16:52.6 & SPM-B\&C & 2015-11-09 & 2$\times$1800 & 3300-7900 & 10.1\\ 
  Swift J0528.1-3933 NE           & 05:28:03.18  &  -39:34:25.7 & SPM-B\&C & 2013-12-04 & 2$\times$1800 & 3300-7900 & 11.2\\
  Swift J0528.1-3933 SW           & 05:28:02.04   & -39:34:44.7  & SPM-B\&C & 2013-12-04 & 1800 & 3300-7900 & 11.2\\
  Swift J0535.2+2830$^{\star}$    & 05:34:57.91 & 28:28:37.3 & TNG-DOLoReS & 2014-01-04 & 2$\times$1200 & 3700-3900 & 16.0\\
  Swift J0550.7-3215 B$^{\star}$  & 05:50:53.62$^{\ddagger}$  &  -32:16:19.0$^{\ddagger}$ & NTT-EFOSC & 2017-03-23 & 3$\times$900 & 4000-8500 & 22.0\\
  Swift J0609.5-6245 B$^{\star}$  & 06:08:38.69 & -62:47:16.2  & NTT-EFOSC & 2017-02-17 & 3$\times$60 & 4000-8500 & 23.5\\
  Swift J0659.3+2406$^{\star}$    & 06:59:10.67  & 24:01:40.3 & TNG-DOLoReS & 2014-01-04 & 2$\times$1200 & 3700-3900 & 15.2\\
  Swift J0701.3-3235$^{\star}$    & 07:04:08.05  &  26:25:03.6 & TNG-DOLoReS & 2014-01-04 & 2$\times$1200 & 3700-3900 & 12.5\\
  Swift J0725.8+3000              & 07:25:37.24  &  29:57:14.8 & Cassini-BFOSC & 2013-04-03 & 2$\times$1800 & 3400-8550 & 13.5\\
  Swift J0733.9+5156$^{\star}$    & 07:33:26.79  & 51:53:56.0 & SPM-B\&C & 2013-03-20 & 2$\times$1800 & 3300-7900 & 11.5\\
  Swift J0755.4+8402$^{\star}$    & 07:58:18.59  &  83:56:37.4 & OAGH-B\&C & 2013-12-07 & 1200 & 3300-7900 & 11.5\\
  Swift J1007.4+6534$^{\star}$    & 10:07:13.67  &  65:35:01.1 & Cassini-BFOSC & 2013-03-19 & 2$\times$1800 & 3400-8550 & 13.5\\
  Swift J1041.4-1740$^{\star}$    & 10:41:01.20  &  -17:34:29.6 & Cassini-BFOSC & 2013-04-03 & 2$\times$1800 & 3400-8550 & 13.5\\
  Swift J1051.2-1704 B            & 10:51:37.46 & -17:07:29.1 & SPM-B\&C & 2013-04-20 & 2$\times$1800 & 3300-7900 & 9.2\\
  Swift J1114.3+2020 SW           & 11:14:02.48 & 20:23:14.3   & Cassini-BFOSC & 2013-03-20 & 2$\times$1800 & 3400-8550 & 13.1\\
  Swift J1114.3+2020 NE           & 11:14:02.76 & 20:23:21.4 & SPM-B\&C & 2015-02-26 & 3$\times$1800 & 3300-7900 & 11.5\\
  Swift J1306.4-4025 B            & 13:07:12.08  & -40:24:27.3 & SPM-B\&C & 2013-04-30  & 2$\times$1800 & 3300-7900 & 11.5\\
  Swift J1512.2-1053 B$^{\star}$  & 15:12:05.06  &  -10:46:35.6 & SPM-B\&C & 2013-04-20 & 3$\times$1800 & 3300-7900 & 14.2\\
  Swift J1643.2+7036              & 16:43:20.25  &  70:37:57.0 & Cassini-BFOSC & 2014-03-25 & 1800 & 3400-8550 & 13.5\\
  Swift J1731.3+1442              & 17:31:13.38  &  14:42:55.8 & SPM-B\&C & 2013-04-22 & 2$\times$1800 & 3300-7900 & 11.5\\
  Swift J1825.7+7215 NE$^{\star}$ & 18:25:29.78 & 72:09:04.1    & TNG-DOLoReS & 2014-05-04 & 2$\times$1200 & 3700-3900 & 14.0\\
  Swift J1825.7+7215 SW$^{\star}$ & 18:25:28.59 & 72:08:59.3  & TNG-DOLoReS & 2014-05-04 & 2$\times$1200 & 3700-3900 & 14.0\\
  Swift J1852.8+3002$^{\star}$    & 18:52:49.43  &  30:04:27.9 & SPM-B\&C & 2013-09-08 & 2$\times$1800 & 3300-7900 & 9.6\\
  Swift J1925.0+5041              & 19:25:02.18  &  50:43:13.9 & SPM-B\&C & 2013-06-12 & 2$\times$1800 & 3300-7900 & 12.5\\
  Swift J2024.0-0246$^{\star}$    & 20:24:02.05 & -02:45:32.6 & SPM-B\&C & 2014-07-03 & 2$\times$1800 & 3300-7900 & 9.8\\
  Swift J2059.6+4301 A$^{\star}$  & 20:59:15.71$^\dagger$  &  43:01:08.5$^\dagger$ & SPM-B\&C & 2014-10-01 & 3$\times$1800 & 3300-7900 & 11.0\\
  Swift J2059.6+4301 B$^{\star}$  & 21:00:01.00  &  43:02:10.9 & SPM-B\&C & 2010-07-15 & 3$\times$1800 & 3300-7900 & 13.0\\
  Swift J2237.0+2543              & 22:36:32.87 &   25:45:47.4 & Cassini-BFOSC & 2009-11-17 & 1800 & 3400-8550 & 13.5\\
  Swift J2242.4-3711              & 22:42:33.38  &  -37:11:07.4 & SPM-B\&C & 2013-12-05 & 2$\times$1200 & 3300-7900 & 11.5\\
  Swift J2352.6-1707$^{\star}$    & 23:52:51.42  &  -17:04:37.2 & SPM-B\&C & 2013-09-06 & 2$\times$1200 & 3300-7900 & 11.0 \\
\hline
  2PBC J0819.2-2508$^{\star}$     & 08:19:16.21 & -25:07:06.21  & NTT-EFOSC & 2017-03-23 & 4$\times$900 & 3800-9200 & 22.0 \\

\hline\hline\end{tabular}
\end{table*}

\section{Optical observations}

In Table 2, we present the list of the 41 optical candidate counterparts for the 35 hard X-ray objects considered in the present work (which were listed in Table 1), and the details about the spectroscopic observations thereof. Column 1 indicates the names of the selected BAT sources. In columns 2 and 3 we report the equatorial celestial coordinates $\alpha$ and $\delta$ at the J2000.0 epoch for the proposed optical counterpart; in column 4 the telescope and instrument used for each observation; in column 5 the observation date in UT; in column 6 the number of frames taken and their exposure time; in column 7 the covered spectral range in \AA; and in column 8 the average wavelength resolution for each spectrum, calculated as the full width at half maximum, in \AA, of the corresponding background oxygen emission sky line at 5575\, \AA. Unless stated otherwise, all coordinates for optical counterparts are taken from the 2MASS catalogue \citep{Skrutskie06}, which provides a positional uncertainty of 0\farcs1; coordinates taken from the USNO-A2.0 
\citep{Monet98} and from the USNO-B1.0 \citep{Monet03} catalogs bear instead a positional accuracy of 0\farcs3. Multiple optical counterparts which correspond to the same BAT source are marked as NE (north-east) and SW (south-west), based on their sky position.

The optical data presented in this work were collected with the following telescopes:

\begin{itemize}

\item the 1.52\,m \emph{Cassini} telescope of the \emph{Osservatorio di Astrofisica e Scienza dello Spazio} (OAS), in Loiano, Italy;

\item the 2.1\,m telescope of the \emph{Observatorio Astron\'omico Nacional} in San Pedro M\'artir (SPM), Mexico;

\item the 2.12\,m telescope at the \emph{Observatorio Astrof\'isico Guillermo Haro} (OAGH) in Sonora, Mexico;

\item the 3.58\,m \emph{Telescopio Nazionale Galileo} (TNG) at the Roque de Los Muchachos Observatory in La Palma, Canary Islands, Spain;

\item the 3.58\,m \emph{New Technology Telescope} (NTT) of the ESO-La Silla Observatory, Chile;

\item and the 1.2\,m \emph{UK Schmidt Telescope} at the \emph{Anglo-Australian Observatory} (AAO) in Coonabarabran, New South Wales, Australia.

\end{itemize}

All observations from the first five entries above were bias corrected, flat-fielded, their background contamination was subtracted, cleaned for cosmic rays and reduced following standard procedures using IRAF\footnote{IRAF is the Image Reduction and Analysis Facility, distributed by the National Optical Astronomy Observatories, which are operated by the Association of Universities for Research in Astronomy, Inc., under cooperative agreement with the National Science Foundation. It is available at {\tt http://iraf.noao.edu}}. All calibration frames were taken on the day preceding or following the observing night, if not on the same day. Wavelength and flux calibrations were performed with the use of lamp acquisitions (always taken immediately before or after a science data frame), and of catalogued spectrophotometric standard stars, respectively. The uncertainty in the wavelength calibration was of 3 \AA\, at most. All spectra taken for a single object were finally stacked together, either by median or sum, to increase the signal-to-noise ratio (S/N).

In the case of the 6dF data, these were downloaded from the 6dFGS archive \citet{Jones09}; since it contains spectra that are not calibrated in flux, we considered the available optical photometric information \citep{Jones05} for their calibration

Since most of the extragalactic sample is of the AGN type, we performed an archival search for radio data, which are useful for verifying both the BL Lac and the (radio loud) Seyfert nature of the selected sources, if any. In Table 3, we show the corresponding fluxes and errors as listed in the FIRST \citep{Becker95} and NVSS \citep{Condon98} surveys. In column 1 we report the source name, in column 2 its radio flux density in the 1.4 GHz band, in column 3 the corresponding survey and in column 4 the our classification for each source. In column 5 we report the \textbf{logarithm of the} radio-loudness parameter \textsc{R}, following \citet{Zhou06}. We also looked for radio data on the SUMMS \citep{Mauch03} and MGPS2 \citep{Murphy07} radio surveys that cover the southern hemisphere, but we did not find any of our sources in those surveys.

\section{Results}

In the following we describe the various methods we used to interpret our data and the results we obtained in each case, for each object class. We show all the optical spectra of our sample in Figures 14 to 19 (emission line galaxies), 20 (elliptical galaxies and blazars) and 21 (galactic sources); objects within each group are ordered by right ascension.

Based on these spectroscopic observations we were able to classify 41 optical objects in total: 36 as extragalactic ones, while the remaining 5 were classified as Galactic sources.

For all extragalactic objects we assumed a cosmology with $\rm{H}_0=70$\,km s$^{-1}$ Mpc, $\Omega_\Lambda=0.7$ and $\Omega_m=0.3$, and all the extragalactic distances presented in this work were obtained using the cosmology calculator of \citet{Wright06}. The spectra corresponding to extragalactic sources were corrected for Galactic absorption, using the estimates given by \citet{Schlafly11} and the Galactic extinction law of \citet{Cardelli89}.

\subsection{Extragalactic sources}

Our sample includes 36 galaxies; 31 of them correspond to various types of emission line galaxies (mainly AGN of Seyfert type). Out of the remaining five, three show typical elliptical galaxy spectra, while the other two show BL Lac-type spectra.  We will discuss these categories separately.

As expected, both BL Lac sources are strong radio emitters, which is a typical behaviour for this kind of objects \citep{Padovani95}. As for the rest of the radio emitting objects in our sample, they are all emission line galaxies, which are mostly classified radio-quiet sources \citep{Rush96}, although this has been put into question \citep{Ho01}. Thus, it is expected that a part of our sample is detectable at the radio frequencies. Indeed, 13 ($\sim$42\%) out of our subsample of 31 emission line galaxies were detected in the radio bands, although none of them is reported to be significantly resolved. We estimated the radio loudness \textbf{parameter} \textsc{R} of these sources, following \citet{Zhou06} and \citet{Tanaka14}, by obtaining the ratio of the 1.4 GHz flux to \textbf{an optical band flux, both restframe, when possible. \textbf{We assumed a flat radio spectrum to obtain the corresponding flux at 1.4 GHz restframe, since the low redshift values present in our sample would provide limited difference in the final result.} We chose to use the B-band flux, as in \citet{Tanaka14}. There is, however, a caveat:} that we cannot differentiate the optical emission arising only from the nucleus. This means that these results are conservative lower limits. We find that only Swift J0725.8+3000 and Swift J0733.9+5156 are radio loud, with \textbf{log(\textsc{R})} parameters 1.5 and 7.4, respectively; while the rest show \textbf{log(\textsc{R})} values below 1.

\begin{table}
\caption{NVSS and FIRST radio detections for sources in our sample. The data were obtained with a 5.0 arcseconds search radius centered on the optical counterpart on both catalogs. In column 1 we report the source name taken from Swift70, in column 2 the integrated radio flux density in the 1.4 GHz band and its error, in column 3 the survey from which the data was taken from, in column 4 our classification of the source and in column 5 \textbf{the logarithm of} its radio-loudness parameter \textsc{R}.}
\begin{center}
\small
\setlength{\tabcolsep}{2pt}

\begin{tabular}{|l|c|l|l|c|}

\hline\hline
Name & Flux$\pm$Err. & Survey & Class & \textbf{log(\textsc{R})} \TBstrut \\
\hline\hline
  & [mJy] &  &  & \\
\hline\hline
Swift J0107.7-1137 B  & 3.5$\pm$0.5   & NVSS  & Seyfert 1.9 & 0.01\\
Swift J0136.5+3906    & 60.0$\pm$1.8  & NVSS  & BL Lac & $>$0.25 \\
Swift J0138.8+2925    & 4.6$\pm$0.4   & NVSS  & Seyfert 2 & 0.13\\
Swift J0222.3+2509    & 5.0$\pm$0.5   & NVSS  & Seyfert 1.9 & 0.06\\
Swift J0528.1-3933 SW & 4.5$\pm$0.5   & NVSS  & Seyfert 1.9 & 0.17 \\
Swift J0725.8+3000    & 146.3$\pm$5.2 & NVSS  & Seyfert 2 & 1.5\\
Swift J0733.9+5156    & 28.09$\pm$0.15 & FIRST & BL Lac & 7.4\\
Swift J1114.3+2020 NE &  8.52$\pm$0.14 & FIRST & LINER & 0.15\\
Swift J1643.2+7036    & 42.7$\pm$1.6   & NVSS  & Seyfert 2 & 0.24\\
Swift J1731.3+1442    & 5.0$\pm$0.5   & NVSS  & Seyfert 1.2 & 0.01\\
Swift J1825.7+7215 NE & 9.9$\pm$0.5   & NVSS  & Seyfert 2 & 0.24\\
Swift J2237.0+2543    & 5.3$\pm$0.4   & NVSS  & Seyfert 1.9 & 0.05\\
Swift J2242.4-3711    & 31.1$\pm$1.0   & NVSS  & Seyfert 2 & 0.1\\
\hline\hline

\end{tabular}

\end{center}
\end{table}

\subsubsection{Emission line AGN}

The majority of our sample consists of galaxies which show evidence of nuclear activity. In particular, we found 29 Seyfert galaxies and one Low ionization Nuclear Emission Region (LINER); we moreover identify one source as a starburst galaxy.

To properly differentiate between type 1 (with broad emission lines) and type 2 
(with narrow emission lines only) AGN, we followed the criteria presented by \citet{Veilleux87}. Then, to determine whether a narrow line source is a Seyfert 2, a type 2 LINER \citep{Heckman80}, or a star forming (i.e. starburst) galaxy \citep{Terlevich85}, we used the ratio diagnostics as shown in \citet{Kauffmann03} and references therein. In Fig. 1 we present the resulting BPT diagram \citep{BPT} for our subsample of narrow emission line galaxies in which we included the separation between different classes as shown both in \citet{Kauffmann03} and \citet{Kewley01}. We did not account for any possible absorption or emission contributions from the host galaxy. For the purposes of our paper, these, if any, could mainly be from $\rm{H}_{\beta}$ and $\rm{H}_{\alpha}$ absorption lines from the stellar population, so the values we report for these two fluxes should be nominally considered as lower limits. We however do not expect any substantial correction due to the relatively low resolution of our spectra. Furthermore, there is one case, that of Swift J1512.2-1053 B, for which the $\rm{H}_{\alpha}$ and $[\rm{NII}]$ lines fell out of the covered spectral range, due to a relatively high value of redshift: this source was thus not included in this diagram. We also discarded one more narrow line source, Swift J0042.9+3016 A, for which the $\rm{H}_{\beta}$ was featured as an absorption line, since this does not allow obtaining lower limit for the $[\rm{O\, III}]/\rm{H}_{\beta}$ flux ratio. In total, the BPT diagram shown in Fig. 1 represents 13 out of our 15 narrow emission line galaxies.

Finally, we followed \citet{Winkler92} to assign a subclass to the Seyfert galaxies of our sample by using the $\rm{H}_{\beta}/[\rm{O\, III}]_{\lambda 5007}$ line flux ratio. Their criteria is as follows:

\begin{itemize}

\item $\rm{H}_{\beta}/[\rm{O\,III}] > 5.0$ : Seyfert 1.0
\item $\rm{H}_{\beta}/[\rm{O\,III}] < 5.0$ : Seyfert 1.2
\item $\rm{H}_{\beta}/[\rm{O\,III}] < 2.0$ : Seyfert 1.5
\item $\rm{H}_{\beta}/[\rm{O\,III}] < 0.3$ and both broad components of $\rm{H}_{\beta}$ and $\rm{H}_{\alpha}$ visible : Seyfert 1.8
\item Broad component of $\rm{H}_{\alpha}$ visible but $\rm{H}_{\beta}$ narrow : Seyfert 1.9
\item Both $\rm{H}_{\beta}$ and $\rm{H}_{\alpha}$ narrow : Seyfert 2

\end{itemize}

The main results obtained from the analysis of the optical spectra of our sample of emission-line galaxies are given in Table 3: we report in column 1 the BAT name, in column 2 its flux in the BAT band as given by Swift70, in column 3 the redshift of the proposed optical counterpart as measured from our optical spectra, in column 4 the resulting luminosity distance, in column 5 the derived luminosity in the BAT band, in columns 6 to 10 the emission lines identified in our spectra for the corresponding optical counterpart alongside their observed and dereddened fluxes and their statistical errors, in column 11 the derived velocity for the broad $\rm{H}_{\beta}$ line (if present), in column 12 the mass of the black hole obtained from such velocity, and finally in column 13 the class we assigned to each source. 

We remark that a preliminar identification of Swift J1852.8+3002 was given in \citet{Halpern13}: both classification (Seyfert 1.9) and redshift ($z=0.057$) provided there are coincident, within errors, with the results we report here. Moreover, the classification of 6 objects of our sample was also reported by \citet{Koss17} (these cases are marked with the symbol $\dagger$ in Table 3): 3 of them are in full agreement with our identifications (Swift J0042.9+3016 A, Swift J0659.3+2406 and Swift J1643.2+7036). For the remaining cases, slight differences between our classification and that in \citet{Koss17} are apparent, possibly due to either long-term spectral variability of the sources, or different signal-to-noise ratio in the analyzed optical spectra, or both: These are Swift J0138.8+2025, Swift J1114.3+2020 NE and Swift J1731.3+1442, classified by \citet{Koss17} as a Seyfert 1.9, a Seyfert 2 and a Seyfert 1.0, respectively.

From our whole subsample of narrow emission-line galaxies, as shown on Fig. 1, only two sources do not fit with the definition of canonical Seyfert 2 AGN emission: Swift J1114.3+2020 NE, which is a type 2 (i.e., with no broad emission lines) LINER, and Swift J0609.5-6245 B, which lies in the starburst region.

In total, our sample consists of 13 type 2, 6 type 1.9, 1 type 1.8, 4 type 1.5, 4 type 1.2 and 1 type 1.0 Seyfert galaxies, plus one LINER and one Starburst as mentioned above.

We also obtained central black hole mass estimations for all our type 1 AGN sources showing a broad $\rm{H}_{\beta}$ emission line. 

In all cases, assuming no local absorption at the AGN site, we measured the full width at half maximum (FWHM) of the $\rm{H}_{\beta}$ emission, along with its flux corrected for the color excess due to absorption along the path of light through the Milky Way with the same procedure we described in Sect. 4: concerning this, in Table 5 we report the intervening Galactic E(B-V) values due to intervening Galactic dust for the broad-line AGN of our sample. Then, we were able to obtain the gas velocity in the broad line region (BLR) using these observables. To this aim, the widths of broad lines were corrected for the broadening due to spectral resolution. Finally, using Eq. (2) in \citet{Wu04}, we estimated the size of the BLR, which we then replaced in Eq. (5) of \citet{Kaspi00} to obtain the mass of the black hole. Given the spectral resolution and S/N of our spectra, we did not include any correction for Balmer absorption from the underlying stellar population, as we deem it not critical at this stage although this approach may reduce the evaluation of the actual intrinsic flux of the emission lines. Therefore, these mass values should be considered as order-of-magnitude estimates for the time being.

An estimate of the intrinsic colour excess E(B-V) as a measure of local extinction in the type 2 AGN was also inferred from the optical spectra when possible (i.e. when all the required emission lines were present in the spectra). We evaluated this parameter by measuring the observed line flux ratio $\rm{H}_{\alpha}/\rm{H}_{\beta}$ and comparing it to the expected value of $(\rm{H}_{\alpha}/\rm{H}_{\beta})_0=2.86$, assumed from the Balmer decremental flux relation for plasmas in standard density and temperature conditions as shown in \citet{Osterbrock89}, and by considering a local dust attenuation law similar to that of the Milky Way \citep{Cardelli89}. We chose not to apply this method to the type 1 Seyferts of our sample, since the disentangling of the $H_\alpha$ broad line and the [N II] narrow emissions is very difficult when dealing with spectra with relatively low resolution as those in the present sample; on the contrary, in type 2 AGN only narrow lines are visible, and thus their fluxes easily measured. Results are given in Table 6, in which we report in column 1 the BAT source name, in column 2 the E(B-V) colour excess due to the dust in the Milky Way (from \citealt{Schlafly11}), and in column 3 the E(B-V) colour excess intrinsic to the AGN, or an upper limit thereof in case only a lower limit for the $\rm{H}_{\beta}$ flux is available. The cases in which no estimate could be given correspond to spectra in which the $\rm{H}_{\beta}$ line is seen in absorption.

In this subsample we found one case, Swift J1825.7+7215 NE, for which the observed $\rm{H}_{\alpha}/\rm{H}_{\beta}$ flux ratio is lower than 2.86: this points to the fact that this object may be a \emph{naked} Seyfert 2 galaxy \citep{Panessa09,Tran11}; that is, an AGN in which the BLR emission is not absorbed locally, but rather it is weak or completely absent. However, the XRT data show a relatively highly absorbed spectrum \citep{Ricci17}. Nevertheless, the multiwavelength information we gathered shows that this object is interacting with Swift J1825.7+7215 SW (see Fig. 12), which lies at a distance of $\sim$6\,arcseconds apart, whereas the XRT source position lies between the two galaxies and has an error circle of $\sim$3.5\,arcseconds (thus positionally consistent with both optical objects). This means that the XRT spectrum is the sum of the emission from the two AGN, and that the absorption seen in it is probably due to the contribution of object SW (a regular Seyfert 2), rather than that of NE (the naked Seyfert 2). 

Furthermore, for all the AGN of our sample the soft X-ray primary continuum flux (0.3-10 keV flux) was analysed in comparison with that of the $[\rm{O\,III}]_{\lambda 5007}$ emission line, which is considered as a proxy of the AGN ionizing power \citep{mulchaey94,bassani99}. For the following analysis, all the soft X-ray fluxes were obtained from the 1SXPS catalogue of Swift/XRT sources. Since this catalogue reports count rates in the XRT band, a physical emission model is necessary to obtain the corresponding fluxes. We did so with the use of the WebPIMMS tool\footnote{WebPIMMS is a Web version of the PIMMS (v4.8f) tool and it is available at {\tt https://heasarc.gsfc.nasa.gov/cgi-bin/Tools/w3pimms/w3pimms.pl}}. In all cases we chose to apply a power-law model with spectral index $\Gamma=1.85$, which is the average value of a large sample of observed emission line AGN for the X-ray continuum, as reported by \citet{Dadina08}. In all cases, the redshift measured from our optical spectra was taken into account when calculating the X-ray fluxes.

The analysis we subsequently performed is twofold, depending on whether the soft X-ray flux is absorption-corrected or not; both approaches are shown in Figure 2. Type 1 galaxies (i.e. Seyferts from types 1 to 1.9) are plotted as blue filled circles and Seyfert 2 galaxies as red filled squares; we have also plotted the naked Seyfert 2 galaxy Swift J1825.7+7215 NE as an empty green triangle.

In a first stage, illustrated in the upper panel of Figure 2, we compared the extinction corrected (intrinsic) $[\rm{O\, III}]_{\lambda 5007}$ flux vs. the observed soft X-ray 0.3-10 keV flux. This kind of comparison provides information on whether the ionizing energy that produces the [OIII] line arrives from the X-ray emission or not \citep{Panessa06,Gomez17}, the level of accretion \citep{Lamastra09} or hints of a Compton thick scenario where X-ray emission is heavily obscured \citep{DeRosa18}. Knowing that the X-ray absorption was not accounted for in these fluxes, the X-ray flux values reported in this panel of Figure 2 should be taken as lower limits. Thus, comparing the extinction corrected $[\rm{O\, III}]_{\lambda 5007}$ flux with the X-ray observed flux may provide hints on whether the X-rays are affected by extinction or not, and to what extent. To better quantify this, we have used the {\it thickness parameter} T \citep{bassani99}.This parameter is defined as the ratio of the observed 2-10 keV flux to the intrinsic (absorption-corrected) $[\rm{O}\,\rm{III}]_{\lambda 5007}$ flux. In order to be consistent with the data, we slightly re-define T by using the flux in the wider soft X-ray band (0.3-10 keV) covered by XRT, and we call it T$^*$. 

This was done in order to look for Compton-thick candidates (CT, defined as sources with intrinsic hydrogen column density $\rm{N}_{\rm{H}}\gtrsim 1.5\times10^{24}\,\rm{cm}^{-2}$; \citealt{turner97,maiolino98}) AGN within our sample. Given the proportionality between E(B-V) and nH values, and the fact that the X-ray fluxes reported here are lower limits, we consider that any sources falling above the T$^*$ limit are CT candidates. In other words, that our optical results suggest these sources \emph{could} be CT sources, and that their X-ray spectra should be studied in more detail to confirm or discard this hypothesis.

The diagram in the upper panel of Fig. 2 is crossed by the $T^*=1$ line, which settles the top-left corner of the diagram as the locus of AGN in CT regime. In this way we identify two CT \emph{candidates}, that is, Swift J1041.4-1740 and Swift J2242.4-3711, which lie well within the CT region in the diagram. Indeed, \citet{Ricci17} reported that the latter showed a nH value of $10^{24.3\pm0.43}$, which is consistent with our suggestion that this source is a CT candidate. However they also reported that Swift J1041.4-1740 showed an nH value of $10^{23.2\pm0.26}$, which contradicts this hypothesis. An X-ray spectral analysis is needed to confirm or discard their CT nature for both.

We also note that the Seyfert 1 and Seyfert 2 galaxies cluster in different parts of the diagram, mainly due to the higher $[\rm{O}\,\rm{III}]_{\lambda 5007}$ intrisic fluxes of the latter. This is due to the X-ray flux being depressed by the column density of the torus, which in the case of type 2 galaxies is expected to be \textbf{observed nearly} edge-on with respect to the line of sight \citep{bassani99}. This pattern is also apparent in the lower panel, where both the $[\rm{O}\,\rm{III}]_{\lambda 5007}$ flux and the soft X-ray 0.3-10 keV flux are absorption-corrected (i.e. the intrinsic fluxes are plotted). This was not observed in previous similar analyses \citep[see, e.g.][]{Berney15,Rojas17} realized over larger samples of sources.

We then attempted to compare the intrinsic X-ray and $[\rm{O}\,\rm{III}]_{\lambda 5007}$ fluxes, as in \citet{Berney15}, in order to shed some light on whether there is evidence for a correlation between these quantities or not. It has long been established \citep{Bohlin78} that there is a proportional relation between the optical colour excess E(B-V) and the hydrogen column density nH, a relation that has been tested several times \citep[see, for example,][and references therein]{Vladilo06,Gudennavar12}. To do so, we derived nH values using the E(B-V)/nH relation provided by \citet{Gudennavar12} and our own measured intrinsic E(B-V) values. In the cases in which these latter were not available, we chose not to take this correction into account. We obtained X-ray fluxes in the same fashion as for the upper panel of Figure 2, but this time extinction-corrected with the derived nH values. Since the nH values derive from optical data, the torus of the AGN may not be properly accounted for: Thus, extinction-corrected X-ray fluxes in this panel should be taken as lower limits. In this panel, the distribution of the Seyfert 2 group is rather homogeneous with respect to their soft X-ray intrinsic flux, but their $[\rm{O}\,\rm{III}]_{\lambda 5007}$ flux covers almost two orders of magnitude. Such a distribution does not deserve special attention, as no trend is apparent from these data.

On the other hand, the type 1 galaxy group seems to follow a linear correlation (in logarithmic terms). By doing a similar analysis with respect to that of \citet{Rojas17}, we have then performed a linear fitting to the data belonging only to the type 1 group, plotted in Fig. 2. The relationship between intrinsic $[\rm{O}\,\rm{III}]_{\lambda 5007}$ and 0.3-10 keV fluxes shows a Pearson correlation coefficient of $R_P=0.42$, which yields a {\it p-value} of 0.1053; therefore, it is only marginally statistically significant ($95\%$ confidence level). The resulting correlation is in any case less significant than that pointed out by \citet{Rojas17} for their Type 1 AGN sample.

In the lower panel of Fig. 2 we also plotted the linear relationship found by \citet{Berney15} for comparison (we here stress the caveat that those authors have used a narrower soft X-ray band, i.e., 2-10 keV). It is shown as the dashed magenta line in the lower panel of Fig. 2. As already noted by \citet{Rojas17} for their sample, we as well highlight that the relation of \citet{Berney15} shows a steeper slope than ours. However, the overall sample is widely spread over the diagram, reinforcing the complexity of this relationship \citep[see][]{heckman05}. We conclude, in agreement with both \citet{Berney15} and \citet{Rojas17}, that there appears to be no more than a marginal correlation between these two quantities.

\begin{figure}
\centering
\includegraphics[scale=0.43,angle=270]{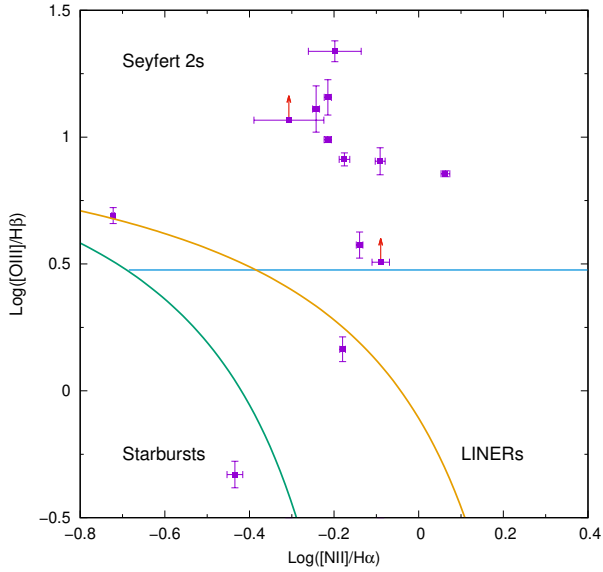}
\caption{BPT diagram ([O III]/H$\beta$ flux ratio vs. [N II]/H$\alpha$ flux ratio) for all the various types of narrow emission line galaxies within the sample presented in this paper. Fluxes are the observed ones (i.e., not corrected for either Galactic or local absorption). Following \citet{Kauffmann03}, the green solid line marks the empirical separation between the dominance of a star formation regime and that of AGN regime, while the light blue solid line shows the separation between AGN and LINER regions. The orange solid line shows again the separation between AGN and LINER regions but following \citet{Kewley01}. All our objects lie within the AGN region, with the exception of two (a starburst galaxy and a LINER; see text), which nonetheless are located close to the loci borders.  There are two upper limits to the H$\beta$ line, which are marked with a red arrow.}
\end{figure}

\begin{figure}
\centering
\includegraphics[scale=0.49,angle=270]{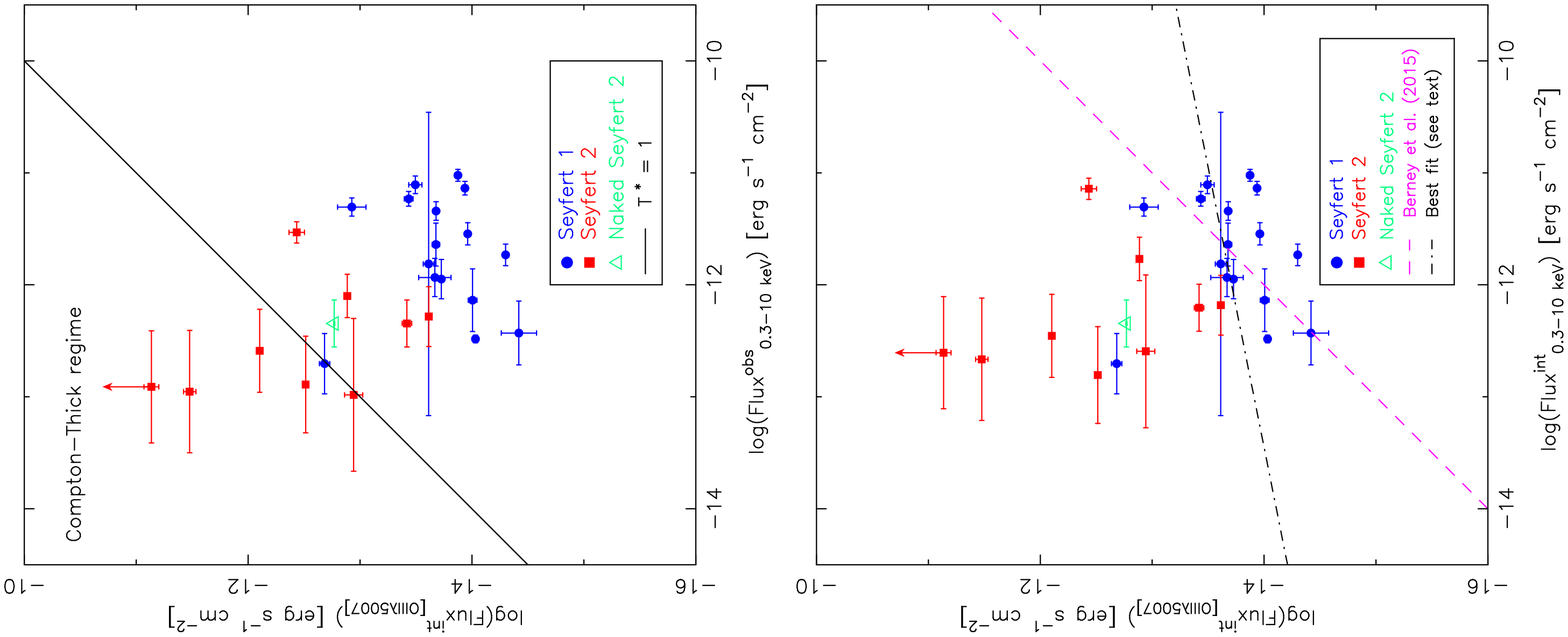}
\caption{Soft X-ray (0.3-10 keV) flux versus [O III] ($\lambda=5007$) emission line flux for a subset of our sample which includes all the emission line galaxies in which the [O III] emission line was observed and for which an XRT counterpart was available in the 1SXPS catalogue. All optical and X-ray fluxes were corrected for the extinction from the Milky Way. In the upper panel we report the observed soft X-ray flux,while in the lower panel we show the X-ray flux corrected for the extinction of its host galaxy}. In both panels, the [O III] emission line flux is the intrinsic one (i.e. extinction-corrected). Type 1 to 1.9 objects are plotted in blue, while type 2 objects are in red. The dashed magenta line represents the correlation reported by \citet{Berney15}, while the dot-dashed black line represents our best fit to the data.
\end{figure}

\begin{sidewaystable*}
\caption{{\tiny Results from X-ray and optical data. In all cases, the reported [O III] and [N II] lines refer to those at 5007\,\AA\ and 6583\,\AA\ restframe, respectively. In some cases
(namely the spectra of Seyfert 1 AGN), the H$\alpha$ and [N II] lines were blended, so we report here the observed total blended flux and EW, along with (if possible) the estimated deblended values for each individual line. The latter are marked with an asterisk. Such deblended flux estimates were obtained through the best possible fitting of 3 gaussian profiles to the observed emission complex. Line doublets reported as blended are marked with a d. Objects for which a classification was given by \citet{Koss17} are marked with a $\dagger$. Flux errors do not include systematic uncertainties. Whenever the $\rm{H}_{\beta}$ line was absent, we derived a minimum value for the AGN emission contribution, which is marked with $\ddagger$.}}

\setlength{\tabcolsep}{2pt}
\small
\centering
\begin{tabular}{|c|c|c|c|c|lrr|c|c|c|}
\hline\hline
  \multicolumn{1}{|c|}{Name} &
\multicolumn{1}{c|}{BAT flux}  & 
  \multicolumn{1}{c|}{z} &
  \multicolumn{1}{c|}{Dist. L.} &
  \multicolumn{1}{c|}{$\mathrm{L}_X$} &
  \multicolumn{1}{l}{Line} &
  \multicolumn{1}{r}{Flux} &
  \multicolumn{1}{r|}{Dered. Flux} &
  \multicolumn{1}{c|}{$V^{FWHM}_{H\beta}$} &
  \multicolumn{1}{c|}{$M_{bh}$} &
  \multicolumn{1}{c|}{Class} \Tstrut \\

  \multicolumn{1}{|c|}{} &
  \multicolumn{1}{c|}{$\left[\frac{erg\,\times\,10^{-12}}{cm^2\,s}\right]$} &
  \multicolumn{1}{c|}{} &
  \multicolumn{1}{c|}{[Mpc]} &
  \multicolumn{1}{c|}{$\left[\frac{erg\,\times\,10^{43}}{s}\right]$} &
  \multicolumn{1}{l}{} &
  \multicolumn{1}{r}{$\left[\frac{erg\,\times\,10^{-14}}{cm^2\,s}\right]$} &
  \multicolumn{1}{r|}{$\left[\frac{erg\,\times\,10^{-14}}{cm^2\,s}\right]$} &
  \multicolumn{1}{c|}{[km/s]} &
  \multicolumn{1}{c|}{[$M_{\odot}\times 10^6$]} &
  \multicolumn{1}{c|}{} \Bstrut \\

\hline\hline

  Swift J0042.9+3016 A $\dagger$ & 7.72 & 0.049$\,\pm\,$0.003 & 218 & 4.4 & [O III] & 2.15$\,\pm\,$0.02 & 2.44$\,\pm\,$0.05 &  &  & Seyfert 2 \Tstrut \\
   &  &  &  &  & H$\alpha$ & 2.575$\,\pm\,$0.002 & 2.81$\,\pm\,$0.03 &  &  & \\
   &  &  &  &  & [N II] & 1.93$\,\pm\,$0.01 & 2.06$\,\pm\,$0.04 &  &  & \\
    \hline
    
  Swift J0052.3-2730 & 8.21 & 0.077$\,\pm\,$0.003 & 349 & 12.0 & H$\beta$ & 0.11$\,\pm\,$0.01 & 0.13$\,\pm\,$0.01 &  &  & Seyfert 2 \\
   &  &  &  &  & [O III] & 1.42$\,\pm\,$0.01 & 1.485$\,\pm\,$0.005 &  &  & \\
   &  &  &  &  & H$\alpha$ & 0.420$\,\pm\,$0.004 & 0.439$\,\pm\,$0.004 &  &  & \\
   &  &  &  &  & [N II] & 0.254$\,\pm\,$0.002 & 0.268$\,\pm\,$0.006 &  &  & \\
    \hline
  Swift J0107.7-1137 A & 10.48 & 0.358$\,\pm\,$0.003 & 1907 & 460 & H$\beta$ & 0.115$\,\pm\,$0.006 & 0.124$\,\pm\,$0.003 & 3550 & 19.9 & Seyfert 1.5\\
   &  &  &  &  & [O III] & 0.092$\,\pm\,$0.001 & 0.161$\,\pm\,$0.001 &  &  & \\
    \hline
  Swift J0107.7-1137 B & 13.51 & 0.047$\,\pm\,$0.004 & 209 & 7.0 & H$\beta$ & 1.5$\,\pm\,$0.2 & 1.8$\,\pm\,$0.1 &  &  & Seyfert 1.9\\
   &  &  &  &  & [O III] & 18.8$\,\pm\,$0.4 & 20.8$\,\pm\,$0.4 &  &  & \\
   &  &  &  &  & H$\alpha$ & 2.70$\,\pm\,$0.06 & 2.88$\,\pm\,$0.02 &  &  & \\
   &  &  &  &  & [N II] & 8.2$\,\pm\,$0.5 & 7.0$\,\pm\,$0.1  &  &  & \\

    \hline

  Swift J0138.8+2925 $\dagger$ & 6.33 & 0.071$\,\pm\,$0.003 & 320 & 7.7 & H$\beta$ & 0.061$\,\pm\,$0.003 & 0.065$\,\pm\,$0.003 &  &  & Seyfert 2\\
   &  &  &  &  & [O III] & 0.49$\,\pm\,$0.01 & 0.492$\,\pm\,$0.004 &  &  & \\
   &  &  &  &  & H$\alpha$ & 0.724$\,\pm\,$0.004 & 0.800$\,\pm\,$0.005 &  &  & \\
   &  &  &  &  & [N II] & 0.587$\,\pm\,$0.006 & 0.66$\,\pm\,$0.01 &  &  & \\
    \hline
  Swift J0222.3+2509 & 15.81 & 0.061$\,\pm\,$0.003 & 273 & 14.1 & H$\beta$ & 0.232$\,\pm\,$0.009 & 0.410$\,\pm\,$0.006 &   &  & Seyfert 1.8 \\
   &  &  &  &  & [O III] & 3.03$\,\pm\,$0.03 & 3.68$\,\pm\,$0.06 &  &  & \\
   &  &  &  &  & H$\alpha$ & 26$\,\pm\,$4 & 34$\,\pm\,$5 &  &  & \\
   &  &  &  &  & [N II] & 1.3$\,\pm\,$0.2 & 1.50$\,\pm\,$0.03 &  &  & \\
    \hline
  Swift J0449.6-5515 & 8.56 & 0.208$\,\pm\,$0.003 & 1024 & 110.0 & H$\beta$ & 6.5$\,\pm\,$0.2 & 6.7$\,\pm\,$0.3 &  &  & Seyfert 2\\
   &  &  &  &  & [O III] & 31.9$\,\pm\,$0.2 & 32.7$\,\pm\,$0.2 &  &  & \\
   &  &  &  &  & H$\alpha$ & 23.04$\,\pm\,$0.07 & 25.10$\,\pm\,$0.08 &  &  & \\
   &  &  &  &  & [N II] & 4.370$\,\pm\,$0.007 & 4.5$\,\pm\,$0.1 &  &  & \\
       \hline
  Swift J0528.1-3933 NE & 10.79 & 0.037$\,\pm\,$0.002 & 163  & 3.4 & H$\alpha$ & 0.465$\,\pm\,$0.002 & 0.516$\,\pm\,$0.001 &  &  & Seyfert 1.9\\
   &  &  &  &  & [N II] & 0.465$\,\pm\,$0.006 & 0.537$\,\pm\,$0.003 &  &  & \\
    \hline
  Swift J0528.1-3933 SW & 10.79 & 0.036$\,\pm\,$0.002 & 158 & 3.2 & H$\beta$ & 0.116$\,\pm\,$0.008 & 0.128$\,\pm\,$0.004 &  &  & Seyfert 1.9 \\
   &  &  &  &  & [O III] & 0.189$\,\pm\,$0.005 & 0.218$\,\pm\,$0.007 &  &  & \\
   &  &  &  &  & H$\alpha$ & 1.02$\,\pm\,$0.06 & 1.4$\,\pm\,$0.2 &  &  & \\
   &  &  &  &  & [N II] & 0.463$\,\pm\,$0.008 & 0.474$\,\pm\,$0.004 &  &  & \\
    \hline
  Swift J0609.5-6245 B & 9.45 & 0.099$\,\pm\,$0.005 & 455 & 23.0 & H$\beta$ & 3.1$\,\pm\,$0.1 & 3.5$\,\pm\,$0.1 &  &   & Starburst \\
   &  &  &  &  & [O III] & 1.45$\,\pm\,$0.06 & 2.1$\,\pm\,$0.1 &  &  & \\
   &  &  &  &  & H$\alpha$ & 10.25$\,\pm\,$0.02 & 11.25$\,\pm\,$0.02 &  &  & \\
   &  &  &  &  & [N II] & 3.77$\,\pm\,$0.07 & 4.05$\,\pm\,$0.09 &  &  & \\
   \hline
  Swift J0659.3+2406 $\dagger$& 12.08 & 0.092$\,\pm\,$0.003 & 421 & 26.1 & H$\beta$ & 0.029$\,\pm\,$0.002 & 0.036$\,\pm\,$0.002 &  &  & Seyfert 2 \\
   &  &  &  &  & [O III] & 0.416$\,\pm\,$0.004 & 0.447$\,\pm\,$0.003 &  &  & \\
   &  &  &  &  & H$\alpha$ & 0.305$\,\pm\,$0.002 & 0.326$\,\pm\,$0.001 &  &  & \\
   &  &  &  &  & [N II] & 0.186$\,\pm\,$0.001 & 0.197$\,\pm\,$0.002 &  &  & \Bstrut \\
  \hline \hline\end{tabular}
\end{sidewaystable*}

\setcounter{table}{3}

\begin{sidewaystable*}
\caption{Continued.}

\setlength{\tabcolsep}{2pt}
\centering
\small
\begin{tabular}{|c|c|c|c|c|lrr|c|c|c|}
\hline\hline

  \multicolumn{1}{|c|}{Name} &
\multicolumn{1}{c|}{BAT flux}  & 
  \multicolumn{1}{c|}{z} &
  \multicolumn{1}{c|}{Dist. L.} &
  \multicolumn{1}{c|}{$\mathrm{L}_X$} &
  \multicolumn{1}{l}{Line} &
  \multicolumn{1}{r}{Flux} &
  \multicolumn{1}{r|}{Dered. Flux} &
  \multicolumn{1}{c|}{$V^{FWHM}_{H\beta}$} &
  \multicolumn{1}{c|}{$M_{bh}$} &
  \multicolumn{1}{c|}{Class} \Tstrut \\

  \multicolumn{1}{|c|}{} &
  \multicolumn{1}{c|}{$\left[\frac{erg\,\times\,10^{-12}}{cm^2\,s}\right]$} &
  \multicolumn{1}{c|}{} &
  \multicolumn{1}{c|}{[Mpc]} &
  \multicolumn{1}{c|}{$\left[\frac{erg\,\times\,10^{43}}{s}\right]$} &
  \multicolumn{1}{l}{} &
  \multicolumn{1}{r}{$\left[\frac{erg\,\times\,10^{-14}}{cm^2\,s}\right]$} &
  \multicolumn{1}{r|}{$\left[\frac{erg\,\times\,10^{-14}}{cm^2\,s}\right]$} &
  \multicolumn{1}{c|}{[km/s]} &
  \multicolumn{1}{c|}{[$M_{\odot}\times 10^6$]} &
  \multicolumn{1}{c|}{} \Bstrut \\

\hline\hline

  Swift J0701.3-3235 & 7.5 & 0.044$\,\pm\,$0.003 & 195 & 3.4 & H$\beta$ & 1.925$\,\pm\,$0.002 & 3.82$\,\pm\,$0.02 & 5880 & 53.1 & Seyfert 1.5 \Tstrut \\
   &  &  &  &  & [O III] & 4.45$\,\pm\,$0.03 & 4.625$\,\pm\,$0.002 &  &  & \\
   &  &  &  &  & H$\alpha$+[N II]d & 10.39$\,\pm\,$0.09 & 12.10$\,\pm\,$0.11 &  &  & \\
   &  &  &  &  & H$\alpha$* & 4.84$\,\pm\,$0.09 & 8.06$\,\pm\,$0.09 &  &  & \\
   &  &  &  &  & [N II]* &   1.04$\,\pm\,$0.02 &   1.51$\,\pm\,$0.02   &  &  & \\
   \hline
   Swift J0725.8+3000 & 9.97 & 0.018$\,\pm\,$0.003 & 78 & 0.7 & H$\beta^{\ddagger}$ & $\leq$0.084 & $\leq$0.101 &  &  & Seyfert 2 \\
   &  &  &  &  & [O III] & 0.270$\,\pm\,$0.03 & 0.38$\,\pm\,$0.03 &  &  & \\
   &  &  &  &  & H$\alpha$ & 1.820$\,\pm\,$0.007 & 2.065$\,\pm\,$0.002 &  &  & \\
   &  &  &  &  & [N II] & 1.48$\,\pm\,$0.03 & 1.71$\,\pm\,$0.02 &  &  & \\
  \hline
  Swift J0755.4+8402 & 8.1 & 0.133$\,\pm\,$0.002 & 626 & 38.2 & H$\beta$ & 4.5$\,\pm\,$0.1 & 9.7$\,\pm\,$0.1 & 4080 & 27.9 & Seyfert 1.2 \\
   &  &  &  &  & [O III] & 1.85$\,\pm\,$0.03 & 2.090$\,\pm\,$0.003 &  &  & \\
  \hline
  Swift J1007.4+6534 & 8.85 & 0.126$\,\pm\,$0.003 & 590 & 37.7 & H$\beta$ & 1.51$\,\pm\,$0.02 & 3.39$\,\pm\,$0.03 & 6060 & 59.6 & Seyfert 1.2\\
   &  &  &  &  & [O III] & 0.929$\,\pm\,$0.002 & 1.045$\,\pm\,$0.002 &  &  & \\
   &  &  &  &  & H$\alpha$+[N II]d & 7.6$\,\pm\,$0.2 & 9.2$\,\pm\,$0.2 &  &  & \\
   &  &  &  &  & H$\alpha$* &   6.1$\,\pm\,$0.1     & 7.3$\,\pm\,$0.1  &  &  & \\
   &  &  &  &  & [N II]* &    1.15$\,\pm\,$0.03    & 1.38$\,\pm\,$0.03   &  &  & \\

    \hline
  Swift J1041.4-1740 & 9.78 & 0.081$\,\pm\,$0.002 & 74 & 0.6 & H$\beta^{\ddagger}$ & $\leq$0.036 & $\leq$0.039 &  &  & Seyfert 2 \\
   &  &  &  &  & [O III] & 0.42$\,\pm\,$0.03 & 0.45$\,\pm\,$0.03 &  &  &  \\
   &  &  &  &  & H$\alpha$ & 0.61$\,\pm\,$0.05 & 0.7$\,\pm\,$0.1 &  &  & \\
   &  &  &  &  & [N II] & 0.301$\,\pm\,$0.003 & 0.340$\,\pm\,$0.001 &  &  & \\
  \hline
  Swift J1051.2-1704 B & 3.72 & 0.017$\,\pm\,$0.003 & 368 & 6.0 & [O III] & 0.433$\,\pm\,$0.009 & 0.509$\,\pm\,$0.004 &  &  & Seyfert 1.9 \\
   &  &  &  &  & H$\alpha$+[N II]d & 0.84$\,\pm\,$0.06 & 1.10$\,\pm\,$0.08 &  &  & \\
   &  &  &  &  & H$\alpha$*&   0.67$\,\pm\,$0.05    &   0.88$\,\pm\,$0.07    &  &  & \\
   &  &  &  &  & [N II]*&   0.126$\,\pm\,$0.009   &   0.17$\,\pm\,$0.01    &  &  & \\

 \hline
  Swift J1114.3+2020 NE $\dagger$& 9.47 & 0.027$\,\pm\,$0.002 & 118 & 1.6 & H$\beta$ & 0.192$\,\pm\,$0.007 & 0.195$\,\pm\,$0.003 &  &  & LINER \\
   &  &  &  &  & [O III] & 0.280$\,\pm\,$0.009 & 0.310$\,\pm\,$0.008 &  &  & \\
   &  &  &  &  & H$\alpha$ & 1.725$\,\pm\,$0.005 & 1.805$\,\pm\,$0.005 &  &  & \\
   &  &  &  &  & [N II] & 1.140$\,\pm\,$0.007 & 1.190$\,\pm\,$0.007 &  &  & \\
 \hline
  Swift J1306.4-4025 B & 16.91 & 0.016$\,\pm\,$0.002 & 69 & 1.0 & H$\beta$ & 2.2$\,\pm\,$0.2 & 2.6$\,\pm\,$0.2 & 1730 & 4.3 & Seyfert 1.5\\
   &  &  &  &  & [O III] & 2.35$\,\pm\,$0.04 & 3.21$\,\pm\,$0.08 &  &  & \\
   &  &  &  &  & H$\alpha$+[N II]d & 22.6$\,\pm\,$0.3 & 27.8$\,\pm\,$0.3 &  &  & \\
      &  &  &  &  & H$\alpha$* & 18.0$\,\pm\,$0.3 &   22.2$\,\pm\,$0.3    &  &  & \\
   &  &  &  &  & [N II]* & 3.38$\,\pm\,$0.05 & 4.16$\,\pm\,$0.05    &  &  & \\

   \hline
  Swift J1512.2-1053 B & 8.31 & 0.164$\,\pm\,$0.003 & 787 & 62.4 & H$\beta$ & 0.057$\,\pm\,$0.004 & 0.078$\,\pm\,$0.002 &  &  & Seyfert 2\\
   &  &  &  &  & [O III] & 0.695$\,\pm\,$0.007 & 0.930$\,\pm\,$0.008 &  &  & \\
  \hline
  Swift J1643.2+7036 $\dagger$& 11.7 & 0.014$\,\pm\,$0.003 & 61 & 0.5 & H$\beta$ & 2.60$\,\pm\,$0.03 & 2.87$\,\pm\,$0.06 &  &  & Seyfert 2\\
   &  &  &  &  & [O III] & 18.65$\,\pm\,$0.05 & 20.85$\,\pm\,$0.05 &  &  & \\
   &  &  &  &  & H$\alpha$ & 8.56$\,\pm\,$0.02 & 9.22$\,\pm\,$0.06 &  &  & \\
   &  &  &  &  & [N II] & 9.9$\,\pm\,$0.1 & 10.85$\,\pm\,$0.08 &  &  & \\
   \hline
  Swift J1731.3+1442 $\dagger$& 11.54 & 0.081$\,\pm\,$0.002 & 368 & 19.3 & H$\beta$ & 22$\,\pm\,$2 & 29$\,\pm\,$3 & 2970 & 14.8 & Seyfert 1.2\\
   &  &  &  &  & [O III] & 9.2$\,\pm\,$0.5 & 11.8$\,\pm\,$0.7 &  &  & \\
   &  &  &  &  & H$\alpha$+[N II]d & 76$\,\pm\,$1 & 91.5$\,\pm\,$0.9 &  &  & \\
      &  &  &  &  & H$\alpha$*   & 61$\,\pm\,$1 &   73.2$\,\pm\,$0.7   &  &  & \\
   &  &  &  &  & [N II]*  &    11.4$\,\pm\,$0.2     &  13.7$\,\pm\,$0.1   &  &  & \Bstrut \\

\hline\hline\end{tabular}

\end{sidewaystable*}

\setcounter{table}{3}
\begin{sidewaystable*}
\caption{Continued.}

\setlength{\tabcolsep}{2pt}
\small
\centering
\begin{tabular}{|c|c|c|c|c|lrr|c|c|c|}
\hline\hline

  \multicolumn{1}{|c|}{Name} &
\multicolumn{1}{c|}{BAT flux}  & 
  \multicolumn{1}{c|}{z} &
  \multicolumn{1}{c|}{Dist. L.} &
  \multicolumn{1}{c|}{$\mathrm{L}_X$} &
  \multicolumn{1}{l}{Line} &
  \multicolumn{1}{r}{Flux} &
  \multicolumn{1}{r|}{Dered. Flux} &
  \multicolumn{1}{c|}{$V^{FWHM}_{H\beta}$} &
  \multicolumn{1}{c|}{$M_{bh}$} &
  \multicolumn{1}{c|}{Class} \Tstrut \\

  \multicolumn{1}{|c|}{} &
  \multicolumn{1}{c|}{$\left[\frac{erg\,\times\,10^{-12}}{cm^2\,s}\right]$} &
  \multicolumn{1}{c|}{} &
  \multicolumn{1}{c|}{[Mpc]} &
  \multicolumn{1}{c|}{$\left[\frac{erg\,\times\,10^{43}}{s}\right]$} &
  \multicolumn{1}{l}{} &
  \multicolumn{1}{r}{$\left[\frac{erg\,\times\,10^{-14}}{cm^2\,s}\right]$} &
  \multicolumn{1}{r|}{$\left[\frac{erg\,\times\,10^{-14}}{cm^2\,s}\right]$} &
  \multicolumn{1}{c|}{[km/s]} &
  \multicolumn{1}{c|}{[$M_{\odot}\times 10^6$]} &
  \multicolumn{1}{c|}{} \Bstrut \\

\hline\hline
  Swift J1825.7+7215 NE & 8.51 & 0.108$\,\pm\,$0.003 & 500 & 25.7 & H$\beta$ & 0.709$\,\pm\,$0.008 & 1.85$\,\pm\,$0.03 &  &  & Seyfert 2\\
   &  &  &  &  & [O III] & 6.93$\,\pm\,$0.01 & 17.00$\,\pm\,$0.03 &  &  & ("Naked")\\
   &  &  &  &  & H$\alpha$ & 1.985$\,\pm\,$0.005 & 3.83$\,\pm\,$0.02 &  &  & \\
   &  &  &  &  & [N II] & 1.21$\,\pm\,$0.01 & 2.26$\,\pm\,$0.02 &  &  & \\
     \hline
  Swift J1825.7+7215 SW & 8.51 & 0.109$\,\pm\,$0.003 & 505 & 26.1 & H$\beta$ & 0.087$\,\pm\,$0.002 & 0.184$\,\pm\,$0.007 &  &  & Seyfert 2\\
   &  &  &  &  & [O III] & 0.711$\,\pm\,$0.008 & 1.73$\,\pm\,$0.02 &  &  & \\
   &  &  &  &  & H$\alpha$ & 0.367$\,\pm\,$0.001 & 0.698$\,\pm\,$0.006 &  &  & \\
   &  &  &  &  & [N II] & 0.245$\,\pm\,$0.003 & 0.46$\,\pm\,$0.01 &  &  &  \\
     \hline
  Swift J1852.8+3002 & 13.64 & 0.058$\,\pm\,$0.002 & 259 & 11.1 & H$\beta$ & 0.05$\,\pm\,$0.002 & 0.101$\,\pm\,$0.003 &  &  & Seyfert 2 \Tstrut \\
   &  &  &  &  & [O III] & 1.09$\,\pm\,$0.01 & 1.88$\,\pm\,$0.02 &  &  & \\
   &  &  &  &  & H$\alpha$ & 0.60$\,\pm\,$0.02 & 1.62$\,\pm\,$0.01 &  &  & \\
   &  &  &  &  & [N II] & 0.38$\,\pm\,$0.02 & 0.449$\,\pm\,$0.005 &  &  & \\
  \hline
  Swift J1925.0+5041 & 10.42 & 0.068$\,\pm\,$0.003 & 306 & 17.3 & H$\beta$ & 10.6$\,\pm\,$0.2 & 12.8$\,\pm\,$0.4 & 10340 & 173.8 & Seyfert 1\\ 
   &  &  &  &  & [O III] & 1.10$\,\pm\,$0.01 & 1.335$\,\pm\,$0.005 &  &  & \\
   &  &  &  &  & H$\alpha$+[N II]d & 19.0$\,\pm\,$0.6 & 25.0$\,\pm\,$0.8 &  &  & \\
   \hline
  Swift J2024.0-0246 & 11.32 & 0.137$\,\pm\,$0.002 & 646 & 57.5 & H$\beta$ & 0.46$\,\pm\,$0.02 & 0.48$\,\pm\,$0.05 & 2940 & 13.4 & Seyfert 1.5 \\
   &  &  &  &  & [O III] & 0.315$\,\pm\,$0.003 & 0.365$\,\pm\,$0.002 &  &  & \\
   &  &  &  &  & H$\alpha$+[N II]d & 1.9$\,\pm\,$0.1 & 2.19$\,\pm\,$0.08 &  &  & \\
   &  &  &  &  & H$\alpha$*&    1.51$\,\pm\,$0.08     &   1.75$\,\pm\,$0.07   &  &  & \\
   &  &  &  &  & [N II]*&   0.28$\,\pm\,$0.02     &   0.39$\,\pm\,$0.01   &  &  & \\
   \hline
  Swift J2059.6+4301 A & 5.49 & 0.065$\,\pm\,$0.002 & 292 & 5.6 & [O III] & 0.064$\,\pm\,$0.002 & 2.10$\,\pm\,$0.03 &  &  & Seyfert 1.9 \\
   &  &  &  &  & H$\alpha$ & 0.080$\,\pm\,$0.001 & 0.920$\,\pm\,$0.003 &  &  & \\
   &  &  &  &  & [N II] & 0.120$\,\pm\,$0.002 & 1.37$\,\pm\,$0.04 &  &  & \\
   \hline
  Swift J2237.0+2543 & 10.23 & 0.025$\,\pm\,$0.003 & 109 & 1.5 & [O III] & 1.01$\,\pm\,$0.03 & 0.99$\,\pm\,$0.02 &  &  & Seyfert 1.9\\
   &  &  &  &  & H$\alpha$ & 0.58$\,\pm\,$0.05 & 0.63$\,\pm\,$0.07 &  &  & \\
   &  &  &  &  & [N II] & 0.383$\,\pm\,$0.001 & 0.410$\,\pm\,$0.003 &  &  & \\
   \hline
  Swift J2242.4-3711 & 10.44 & 0.028$\,\pm\,$0.002 & 123 & 1.9 & H$\beta$ & 0.82$\,\pm\,$0.04 & 0.86$\,\pm\,$0.02 &  &  & Seyfert 2 \\
   &  &  &  &  & [O III] & 3.08$\,\pm\,$0.05 & 3.50$\,\pm\,$0.06 &  &  & \\
   &  &  &  &  & H$\alpha$ & 11.05$\,\pm\,$0.02 & 11.10$\,\pm\,$0.10 &  &  & \\
   &  &  &  &  & [N II] & 8.01$\,\pm\,$0.06 & 8.1$\,\pm\,$0.1 &  &  & \\
   \hline
  Swift J2352.6-1707 & 11.72 & 0.055$\,\pm\,$0.002 & 245 & 8.4 & H$\beta$ & 5.19$\,\pm\,$0.02 & 5.3$\,\pm\,$0.1 & 8100 & 102.9 & Seyfert 1.2\\
   &  &  &  &  & [O III] & 1.12$\,\pm\,$0.02 & 1.155$\,\pm\,$0.005 &  &  & \\
   &  &  &  &  & H$\alpha$+[N II]d & 30.5$\,\pm\,$0.7 & 34.1$\,\pm\,$0.9      &  &  & \\
   &  &  &  &  & H$\alpha$*& 24.4$\,\pm\,$0.5 & 27.2$\,\pm\,$0.7    &  &  & \\
   &  &  &  &  & [N II]*&   4.6$\,\pm\,$0.1    &   5.1$\,\pm\,$0.1    &  &  & \Bstrut \\
   
\hline\hline\end{tabular}

\end{sidewaystable*}

\begin{table}
\caption{Line-of-sight extinction values within the Milky Way (MW), obtained from the maps of \citet{Schlafly11}, for the broad emission line galaxies of our sample.}
\begin{center}
\begin{tabular}{|c|c|}

\hline\hline
Name & E(B-V)$_{MW}$  \TBstrut \\
\hline\hline
Swift J0107.7-1137 A &  0.03 \\
Swift J0107.7-1137 B & 0.02    \\
Swift J0222.3+2509 &  0.06   \\
Swift J0528.1-3933 NE &  0.02   \\
Swift J0528.1-3933 SW &  0.02   \\
Swift J0609.5-6245 &   0.05  \\
Swift J0701.3-3235 &   0.01  \\
Swift J0755.4+8402 &   0.05  \\
Swift J1007.4+6534 &   0.05  \\
Swift J1051.2-1704 B &   0.02  \\
Swift J1306.4-4025 B &  0.09   \\
Swift J1643.2+7036 &  0.03  \\
Swift J1731.3+1442 &  0.08  \\
Swift J1925.0+5041 &  0.08  \\
Swift J2024.0-0246 &  0.05  \\
Swift J2059.6+4301 A & 1.17   \\ 
Swift J2237.0+2543 &   0.04  \\
Swift J2352.6-1707 &   0.02 \\
\hline\hline

\end{tabular}

\end{center}
\end{table}

\begin{table}
\setlength{\tabcolsep}{2pt}
\small
\centering
\caption{Extinction values, both from the Milky Way (MW; from \citet{Schlafly11}) and intrinsic to the source (AGN), for the narrow emission line galaxies in our sample. The intrinsic values were derived when both H$\beta$ and H$\alpha$ are detected in the corresponding spectrum, and should be taken as broad estimates given the limitations of the applied method.}\begin{center}
\begin{tabular}{|c|c|c|}

\hline\hline
Name & E(B-V)$_{MW}$ & E(B-V)$_{AGN}$ \TBstrut \\
\hline\hline

Swift J0042.9+3016 A & 0.05 & - \,$\left( \ast \right)$ \\
Swift J0052.3-2730   & 0.01 & 0.17 \\
Swift J0138.8+2925   & 0.04 & 1.45 \\
Swift J0449.6-5515   & 0.01 & 0.28 \\
Swift J0659.3+2406   & 0.06 & 1.17 \\
Swift J0725.8+3000   & 0.05 & - \,$\left( \ast \right)$ \\
Swift J1041.4-1740   & 0.05 & $>$\,2.53 \,$\left( \dagger \right)$ \\
Swift J1114.3+2020 NE & 0.02 & 1.20 \\
Swift J1512.2-1053 B & 0.09 & - \,$\left( \ddagger \right)$ \\
Swift J1643.2+7036   & 0.04 & 0.12 \\
Swift J1825.7+7215 NE & 0.05 & 0 \,$\left( \diamond \right)$ \\
Swift J1825.7+7215 SW & 0.05 & 0.29 \\
Swift J1852.8+3002 &  0.18  & 1.74 \\
Swift J2242.4-3711   & 0.01  & 1.52 \\
\hline\hline

\end{tabular}
\end{center}

{\small {\bf Notes:} $\left( \ast \right)$: The H$\beta$ line was detected in absorption. $\left( \dagger \right)$: The H$\beta$ emission was not detected, so a minimum value for E(B-V)$_{AGN}$ was estimated from the upper limit of the line flux. $\left( \ddagger \right)$: The H$\alpha$ line is out of range. $\left( \diamond \right)$: The observed Balmer decrement ratio was lower than 2.86, so we consider this source as a "Naked" Seyfert 2 candidate.}
\end{table}

\subsubsection{Elliptical galaxies (XBONGs)}

Swift J0550.7-3215 B, Swift J0107.7-1137 C and Swift J1114.3+2020 SW show spectra typical of elliptical galaxies (see Fig. 20). In all cases, the optical continuum can be explained by the superposition of the contributions of several older, low mass star populations in an overall evolved system. This means that the continuum is thermal, brighter at the red end of the spectrum, and with only absorption lines typical of those kind of stars.  

We however calculate the break contrast at 4000\,\AA\ to assess the possibility of hidden non-thermal emission contributing to the overall spectrum, as one would expect in the case of an AGN of BL Lac type caught during a low-emission phase. To do this, we followed the procedure given by \citet{Laurent98}, that is

\begin{equation*}
Br_{4000}=\frac{\mathrm{f}^{+}-\mathrm{f}^{-}}{\mathrm{f}^{+}}
\end{equation*}

\noindent
where $\mathrm{f}^{+}$ is the integrated spectral flux in the range 4050-4250\,\AA\ and $\mathrm{f}^{-}$ the one in the range 3750-3950\,\AA. It is expected that the BL Lac non-thermal emission (if present) significantly contributes more at the blue end of the spectrum than at the red end, thus making the break contrast value smaller. Following this procedure, a value of $Br_{4000}$ above 0.4 means the object is classified as a normal galaxy while, if below 0.25, then it implies a BL Lac classification. Values between these two limits are to be further scrutinized, and the corresponding sources are retained under the ``BL Lac candidate" category \citep{Stocke1991}. All the three above mentioned galaxies present a value of $Br_{4000}$ larger than 0.4. We therefore conclude that these objects are elliptical galaxies. Given the positional association with soft X-ray emission for candidate counterparts Swift J0107.7-1137 C and Swift J1114.3+2020 SW, we can classify them as X-ray bright, optically normal galaxies (XBONGs; \citeauthor{Comastri02} \citeyear{Comastri02}). The candidate counterpart Swift J0550.7-3215 B, instead, does not show any signs of soft X-ray emission in the available data: this means it is highly improbable that this source contributes to the hard X-ray emission observed with BAT and labeled as Swift J0550.7-3215, being it a normal elliptical galaxy with no X-ray activity. We moreover note that the position of this optical object lies outside of the Swift70 BAT error circle for this high-energy source. We thus conclude that the optical counterpart labeled as Swift J0550.7-3215 A in the Swift70 catalogue is the actual responsible for this BAT detection, since it has been identified with PKS 0548-32, a known very luminous BL Lac object.

Our results concerning these three galaxies are summarized in Table 7, in which we report the BAT source name in column 1, the redshift value measured from our spectra for their proposed optical counterparts in column 2, the $Br_{4000}$ derived value in column 3 and its corresponding error, the resulting luminosity distance in column 4, the measured BAT flux in column 5 and the resulting X-ray luminosity in the BAT band (14-195 keV) in column 6.

\begin{table*}
\caption{Main information for the two XBONGS and the only normal elliptical galaxy of our sample.}
\begin{center}
\small
\begin{tabular}
{|c|c|c|c|c|c|}

\hline\hline
\multicolumn{1}{|c|}{Name} & 
\multicolumn{1}{c|}{z} & 
\multicolumn{1}{c|}{Br$_{\lambda 4000}$} &
\multicolumn{1}{c|}{$d_{\rm L}$} &
\multicolumn{1}{c|}{BAT flux}  & 
\multicolumn{1}{c|}{L$_X$} \Tstrut \\

\multicolumn{1}{|c|}{}&  
\multicolumn{1}{c|}{}& 
\multicolumn{1}{c|}{}&
\multicolumn{1}{c|}{[Mpc]} &
\multicolumn{1}{c|}{$\left[erg\,cm^{-2}\,s^{-1}\,\times\,10^{-12}\right]$} &
\multicolumn{1}{c|}{$\left[erg\,s^{-1}\,\times\,10^{44}\right]$}  \Bstrut \\
\hline\hline

Swift J1114.3+2020 SW & 0.025$\,\pm\,$0.003 & 0.40$\,\pm\,$0.12 & 109 & 9.47 & 0.1 \Tstrut \\
Swift J0107.7-1137 C & 0.205$\,\pm\,$0.002 & 0.49$\,\pm\,$0.14 & 1008 & 4.09 & 5.0 \\
\hline\hline
Swift J0550.7-3215 B & 0.183$\,\pm\,$0.004 & 0.41$\,\pm\,$0.23 & 888 & - & - \Bstrut \\
\hline\hline

\end{tabular}
\end{center}
\end{table*}

\subsubsection{BL Lacs}

One further object presents an optical spectrum similar to an elliptical galaxy, i.e. Swift J0733.9+5156, at redshift $z=0.064\pm0.002$. As we obtained a contrast break value of $Br_{4000}=0.23\,\pm\,0.03$ from it, we classified it as a BL Lac object observed in a low activity state.

The value for this source means that there is evidence of a blue contribution to its continuum that cannot be attributed to a regular elliptical galaxy nature. This indicates that, even though its spectrum is similar to an elliptical galaxy, we classify it as a BL Lac object which was observed in a low activity state. During such phases, the jet of a BL Lac object does not hide the host galaxy contribution but manifests itself as a blue excess \citep{Laurent98,Raiteri16}. This allows for the detection of lines to pinpoint a redshift, such as in the case of Swift J0733.9+5156.

Yet, as the signal-to-noise ratio for the spectrum of this object is of $\sim$10 across its blue side, further information (namely, better quality optical spectroscopy) is needed to definitely confirm this classification. As there are no emission lines detected in the optical spectrum the criterion of \citet{Stickel91}, which implies the search for features with restframe equivalent width (EW) below 5\AA, cannot be used. Thus, the analysis of the available multiwavelength information becomes mandatory to confirm our proposed classification. Indeed, this object has been classified as a `beamed AGN' in the Swift105 catalogue, which according to the definition reported in Swift105 means that it is an AGN but not a Seyfert \emph{`based on the presence and shape of optical emission lines in the literature'' \citep{Oh18}}. Moreover, the positions of its X-ray, optical and radio counterparts are consistent with one another and all lie within the positional uncertainty of the $\gamma$-ray source 3FGL\,J0733.5+5153 \citep{Acero15}, which was classified as a BCU (Blazar Candidate of Uncertain type). This means that it shows typical general characteristics of a blazar in the GeV energy band. It is also positionally consistent with a source detected at very high energies (i.e. in the TeV band, as listed in the \emph{TeVCAT}, \citealt{WakelyHoran08}). So we conclude this object is a blazar of the BL Lac kind, based on its optical spectrum and the multiwavelength evidence discussed here.

Finally, the remaining extragalactic, non-emission line source of our sample (Swift J0136.5+3906) was classified as a BL Lac in Swift70, but as a `beamed AGN' in Swift105, and no optical spectrum was available for it in the literature before our observations. \citet{Piranomonte07} opt for the former classification as well. It is, moreover, listed as a $\gamma$-ray source in the Third Fermi catalogue \citep{Acero15}, as well as a BL Lac in the 5th edition of the Roma-BZCAT catalogue \citep{Massaro15}. It was also detected in the TeV band \citep{WakelyHoran08}.

Its optical spectrum indeed presents a featureless, blue shape typical of blazars of the BL Lac type (see Fig. 20), with neither intrinsic emission nor absorption lines are present in its optical spectrum: this is a key characteristic of the spectra of AGN with collimated jets oriented along the observer's line of sight. Its optical spectral shape, with a marked rise in the flux towards shorter wavelengths, can be explained by a non-thermal emission process, best described with a power law. Unfortunately, the lack of spectral features does not allow us to estimate a redshift for the source, or an upper limit thereof.

\subsection{Galactic sources}

Out of the 5 galactic sources we identified, 4 are Cataclysmic Variables (CVs), while the remaining one is a low mass X-ray binary (LMXB). We will discuss the latter separately. As with the extragalactic subsample, we searched for radio data available online in the literature for these sources, but in this case we did not find anything. This kind of source is known to be relatively faint when detected in radio \citep{Mason04}. Besides, they are not regular emitters, since the radio emission is usually associated only to their outburst phase \citep{Coppejans16}, or in the propeller phase for LMXBs \citep{Archibald09}. Thus, we do not expect for them to be listed in radio catalogs.

\subsubsection{CVs}

Swift J0503.7-2819, Swift J0535.2+2830, PBC J0819.2-2508 and Swift J2059.6+4301B show optical spectra typical of CVs (see Fig. 21). They indeed display Balmer, He I and He II emission lines at zero redshift, meaning these objects are located within our Galaxy. This combination of hydrogen and helium emission lines superimposed on an intrinsically blue optical continuum is typical of spectra in disk-accreting compact sources; specifically, those with a white dwarf in accretion in a close binary system. The results from the analysis of our CV spectra, together with the relevant physical parameters which we could obtain from their study, are shown in Table 8. All distances reported in the table were derived assuming an R-band absolute magnitude of 8.1 following \citet{Warner95}.

The flux of the Balmer emission lines can also be used to estimate the Galactic reddening towards each object. Here however, differently from the case of AGN, the typically observed ($\rm{H}_{\alpha}/\rm{H}_{\beta})_0$ line ratio for unabsorbed CVs is $\sim$1 \citep[e.g.][]{Vojkhanskaja17}, because of photon trapping effects \citep[see][and references therein]{Elitzur83}. 

In the case of Swift J0535.2+2830, both Balmer decrements of 1.0 and 2.86 (which means with and without assuming photon trapping effects in the optical emission mechanism) are valid, since the observed $\rm{H}_{\alpha}/\rm{H}_{\beta}$ ratio is 3.66. The derived extinction assuming photon trapping is of $\rm{A}_{\rm{V}}=4.0$, which is consistent with the value obtained from the Galactic dust maps of \citet{Schlafly11} ($\rm{A}_{\rm{V}}=3.9$). This would mean this object lies outside (or at the outer border of) the Galaxy, since those maps measure the whole amount of gas and dust along a given line of sight within the Milky Way. Thus, we conclude the photon trapping effect hypothesis is hardly applicable for this source, meaning in turn that a more likely value for the actual Balmer decrement is 2.86; the results derived under this assumption are reported in Table 8.

For Swift J0503.7-2819 and Swift J2059.6+4301 B, the observed ratios are such that $1.0 \le \rm{H}_{\alpha}/\rm{H}_{\beta} \le 2.86$, which means that photon trapping effects must be considered. Thus, in these cases the reported extinction value was computed assuming a decrement of 1.0. Lastly, Swift J0535.2+2830 shows a line ratio below 1.0, meaning there is no appreciable amount of extinction along the line of sight of the source to be taken into account.

We can moreover explore the magnetic nature of these CVs via the EWs of the $\rm{H}_{\beta}$ and $\rm{HeII}_{4686\AA}$ lines. In particular, we can study the possibility that the white dwarf is a magnetised star, thus meaning that the system could be classified as a polar or intermediate polar CV \citep[see][and references therein]{Warner95}. Within our sample, one CV (PBC J0819.2-2508) presents an EW ratio $\rm{HeII}/\rm{H}_{\beta}$ below 0.5, suggesting it is a CV without significant magnetic activity. Two cases (Swift J0503.7-2819 and Swift J0535.2+2830) show values between 0.5 and 1 for this parameter, meaning that they possibly are Intermediate Polar CVs. Finally, Swift J2059.6+4301 B presents a $\rm{HeII}/\rm{H}_{\beta}$ ratio of $\sim$1.5, indicating that it likely belongs to the polar subclass of magnetic CVs.

\subsubsection{The case of LMXB Swift J0428.2-6704 A}

The spectrum we obtained for Swift J0428.2-6704 A (see Fig. 21) shows a smooth red continuum and only two lines, both of the Balmer series: a broad, bright $\rm{H}\alpha$ emission line, and a very faint $\rm{H}\beta$ emission line. There are neither other emission lines present, nor intrinsic absorption ones. As already mentioned, the continuum presents a rather smooth, curved shape, and it appears much brighter in the red section than in the blue one. This spectral shape suggests that the object is a Galactic accreting X-ray binary, most likely a LMXB. There is also one NuSTAR \citep{Harrison13} observation available in the archive at the ASI-SSDC web site\footnote{{\tt http://www.asdc.asi.it}}, in which this source is detected above 10 keV. Results from our spectrum are reported in the last line of Table 8.

Indeed, \citet{Strader16} found an eclipsing low-mass X-ray binary associated with the $\gamma$-ray source 3FGL\,J0427.9-6704, detected by the LAT instrument onboard the Fermi satellite. This X-ray binary corresponds to the source detected both by INTEGRAL (as IGR\,J04288-6702; \citealt[]{Grebenev13}) and by Swift70. Although there are actually two Swift detections within the Fermi positional uncertainty error ellipse \citep[see also][]{Parisi09}, those authors were able to distinguish which one of the two corresponds to the $\gamma$-ray emitter. They identified it as Swift\,J0428.2-6704 A, the same object dealt within this subsection, and they classified it as a LMXB hosting an accreting neutron star, possibly a transitional millisecond X-ray pulsar. The other source detected by Swift70, Swift\,J0428.2-6704 B, is a type 1.2 Seyfert galaxy \citep{Parisi14}. 

Our spectrum for Swift\,J0428.2-6704A is actually different from those illustrated by \citet{Strader16} acquired between August 2015 and March 2016. Those spectra are all brighter towards shorter wavelengths, thus definitely bluer than ours, with bright, broad, double-peaked emission lines and with occasional absorption lines. 

In comparison, as stated above, our spectrum shows no absorption lines, only two (Balmer) emission lines, and an overall red shape. We do not detect double-peaked lines, but it is worth noticing our spectral resolution is coarser than that of the spectra in \citet{Strader16}. The differences between the two data sets can possibly be interpreted as due to a decrement in the X-ray activity of the source between March 2016 and February 2017, meaning that the accretion rate was lower during our observations and that we were testifying a low emission state, or the beginning of a transition to the radiopulsar state, in this LMXB (see the discussion in \citealt[]{Strader16}).

\section{Cases of interest}

As stated previously, the majority of our sample (23 cases) corresponds to single BAT detections each associated with a single X-ray and optical counterpart candidate, which is the most straightforward way to identify the actual nature of the source. In Fig. 3, we show an optical image taken from the DSS of the field corresponding to Swift J0222.3+2509 as an example of this statement.

On the other hand, there are a number of optical sources in this sample that are not the univocally proposed counterpart to their respective BAT detections. In total, as stated in Sect. 2, the BAT objects from our sample with multiple optical candidate counterparts are 12. Out of them we already discussed the cases of Swift J0550.7-3215 B (Sect. 4.1.2) and Swift J0428.2-6704 A (Sect. 4.2.2.). In the following we summarize our work aimed at disentangling the actual counterpart of the remaining 10 cases of our sample which show the identification issues described here.

\subsection{Multiple X-ray sources}

\textbf{Swift J0042.9+3016:} The Swift70 catalogue reports a BAT detection, Swift J0042.9+3016, with two possible optical counterparts, A and B. While B is already classified as a Seyfert 1 galaxy in Swift70, the nature of A was still unknown, since it was simply labeled as a source of extragalactic nature. From our spectrum, we indeed identify object A as a Seyfert 2 galaxy. This is in agreement with the classification reported for this same object in Swift105. Interestingly, although Swift105 does provide two BAT localizations for this hard X-ray source, the optical counterpart of object A does not lie within the positional uncertainty of the corresponding BAT detection, while that of object B does. Nevertheless, the optical counterpart of source A lies within the positional uncertainty of the BAT detection provided in Swift70, while that of B does not (see Fig. 4). The XRT data shows that source B is brighter than A, and that source A is not detected above 6 keV. Given this scenario, it is most probable that source B contributes more than A to the BAT detection labeled as Swift J0042.9+3016, if not responsible for the whole hard X-ray emission observed above 15 keV.

\textbf{Swift J0107.7-1137:} The case of Swift J0107.7-1137 A, B and C is complex. In Fig. 5, we show an optical image of the region taken from the DSS-II-Red survey, in which we superimposed the different detections as reported by Swift70 (one) and Swift105 (three), together with the 3 soft X-ray/optical counterparts (highlighted in magenta) indicated in Swift105, with their optical spectra reported in the present work. The BAT detections from each catalogue are given with their corresponding 90\% confidence error radii. From Fig. 5 one can note that the three BAT positions from Swift105 are basically consistent with one another and with the one of Swift70, so that they may well represent one single hard X-ray source. In any case, as it is evident, the optical source associated with Swift J0107.7-1137 C lies well outside any of the BAT detections of both catalogs. Indeed, the only optical object mentioned in Swift105 which lies within all of the four error circles is the one associated with Swift J0107.7-1137 B. After analysing their optical spectra, we identify object A as a Seyfert 1.5 object (see Fig. 14), object B as a Seyfert 1.9 (see Fig. 14), and object C as an elliptical galaxy (see Fig. 20). 

However, when we searched for soft X-ray emission from these sources, we found that object A lacks an XRT counterpart, whereas both objects B and C appreciably emit X-rays. A quick look at their XRT spectra reveals that the spectral index of source B ($\Gamma=1.18^{+0.30}_{-0.25}$) corresponds to a harder spectrum than that of source C ($\Gamma=1.57^{+0.21}_{-0.20}$), albeit marginally. To confirm this, we inspected the NuSTAR archive, in which we found an observation of this field in the 3-79 keV band. Again, source A does not show any X-ray emission in this band. Both objects B and C are well detected in the total energy range; however, source C is not detected above 15 keV, whereas B still is (with S/N of $\sim$4). This confirms that the emission from source B is harder than that from C. So, we deem object B is the only responsible for the whole hard X-ray emission detected with BAT, since it lies within all the BAT error circles associated with Swift J0107.7-1137, it shows a bright soft X-ray counterpart with a hard spectral index, and it displays the optical spectrum typical of an AGN.

\textbf{Swift J0609.5-6245:} The Swift70 catalogue does not provide information on the nature of the two proposed counterparts for Swift J0609.5-6245, labeled as objects A and B. Instead, the Swift105 catalogue reports that object A is a Seyfert 1.9 galaxy, while no information is given about the nature of object B. From our spectrum, we classify Swift J0609.5-6245 B as a Starburst galaxy (see Fig. 15). Both objects present XRT counterparts in the 1SXPS catalogue. While Swift70 reports only one position for this source, Swift105 reports two BAT objects in this field, one for each of the proposed optical counterparts (see Fig. 6). Interestingly, object A does not lie within the error circles reported by Swift105, but is located within the one reported by Swift70. Source B, on the other hand, lies within both error circles reported in Swift105 but outside the Swift70 detection, at a distance of $\sim$9 arcmin from the detection centroid.
It is therefore possible that both objects contribute to the total emission observed with BAT (which is also marginally consistent with being due to one single hard X-ray source), since no evidence is strong enough to rule out any of them.

\textbf{Swift J1051.2-1704:} This BAT source presents two optical candidate counterparts. According to Swift70, source A is a Seyfert 2 galaxy lying within the positional uncertainty circle given in that catalogue \citep{Koss17}. We identify source B as a Seyfert 1.9 galaxy (see Fig. 16) which lies $\sim$7\,arcmin from the centroid of the detection given by Swift70. On the other hand, Swift105 provides two separate detections, which do not superimpose to each other. Each optical source lies within its own corresponding associated Swift105 BAT detection, likely meaning that both are contributing to the total hard X-ray emission (see Fig. 7).

\textbf{Swift J1306.5-4025:} Swift J1306.4-4025 A and B are both Seyfert type galaxies, B being identified in this work as a Seyfert 1.5 (see Fig. 17). The source marked as A, on the other hand, is classified as a 1.2 Seyfert galaxy in Swift70 and as a 1.5 Seyfert galaxy in Swift105. Moreover, while the Swift70 catalogue does not report any identification for object B other than "Galaxy", the Swift105 catalogue claims object B is a Galaxy cluster - a possibility which we rule out from our optical spectrum. Both objects A and B lie outside the Swift70 positional uncertainty circle, A being closer than B to the BAT centroid ($\sim$4 versus $\sim$7 arcminutes; see Fig. 8).
However, Swift105 does provide two different detections, albeit partially superimposed. In this case, B lies within its corresponding positional uncertainty circle, while A lies again outside its own, but at short distance ($\sim$2.2 arcmin from the centroid). NuSTAR observations reveal that both objects emit in hard X-rays, with B presenting a slightly harder spectrum: it is detected with a S/N of $\sim$5 in the 10-15 keV band, in comparison to a S/N of $\sim$3 for source A. There is instead no detection for any of the two sources in the NuSTAR data above 15 keV. In all, it is probable that the optical object labeled as A is indeed the optical counterpart of Swift J1306.4-4025, while more information is needed to rule out object B as a
hard X-ray emitter contributing to the BAT detection(s).

\textbf{Swift J1512.2-1053:} We provide in this work an optical spectrum for object Swift J1512.2-1053 B, which is a Seyfert 2 galaxy (see Fig. 17). There is no information available for the nature of object A, either in the Swift70 or in the Swift105 catalogs, the latter only claiming object A is a "beamed AGN". As Seyfert 2 galaxies are well known emitters in the whole X-ray band of the electromagnetic spectrum, object B is probably responsible for at least part of the emission detected with BAT. There are, however, counterparts associated with object A detected at longer wavelengths (see Fig. 9 for an optical sky chart of this case): a radio source detected by NVSS \citep{Condon98} and an infrared source detected by WISE \citep{Wright10}, along with a soft X-ray object observed with XRT. These counterparts, positionally consistent with one another at subarcsecond accuracy, do not coincide with the only nearby optical source, which lies at a distance of $\sim$7\,arcsec from them and which we identify as an M0 type main sequence star with no peculiarities according to a spectrum obtained on June 2012 at SPM. The soft X-ray, infrared and radio counterparts above were moreover associated with a $\gamma$-ray source detected with Fermi-LAT, and have been proposed as being due to a blazar-type AGN \citep{DAbrusco14}. If this is the case, then object A can also be contributing to the overall hard X-ray emission seen with BAT. This is also the most probable case, since both objects are detected by NuSTAR above 10 keV, with object B showing emission up to 79 keV. More information is thus needed to confirm or rule out this possibility, however: for example, an infrared spectrum for this object would provide hints on the nature of its emission.

\textbf{Swift J2059.6+4301:} The case of Swift J2059.6+4301 A and B is straightforward: we classified both objects in this work. The candidate optical counterpart A shows the spectrum of a Seyfert 1.9 object (see Fig. 18), while the proposed optical counterpart labeled B is a cataclysmic variable (see Fig. 21). Both objects, thus, are of X-ray emitting nature. Given that both positions lie within the error circle reported by the Swift70 catalogue for the hard X-ray associated detection (see Fig. 10), it is probable that both objects contribute to the total BAT emission named Swift J2059.6+4301, although it must be noted that source A is detected with XRT with a S/N of 3.5 in the 6-10 keV range, whereas B is not.

\subsection{Interacting galaxies}

We also found a few cases of X-ray emitting extragalactic sources that are apparently undergoing interactions. There is much evidence that interaction between galaxies can trigger AGN-type activity, although this is still under debate \citep[e.g.][]{Manzer14,Cotini13}. In our sample we found at least two pairs of Seyfert galaxies closely interacting with each other and associated with a soft X-ray emission: Swift J0528.1-3933 NE and SW (see Fig. 11), and Swift J1825.7+7215 NE and SW (see Fig. 12). This is not only apparent from the images of the corresponding fields, but it is also confirmed from their measured redshifts (see Fig. 15 and 17): their values, for each pair of galaxies, differ by $\Delta z=0.001$, which is within the error on their measurement. In both cases the disk structure is visible, with appreciable tidal tails and extensions. These two systems are thus cases of major mergers \citep{Byrd87}. Both galaxies in the Swift J0528.1-3933 system are type 1.9 Seyferts (although we stress that the soft X-rays detected with XMM-Newton are positionally consistent with the SW galaxy only), whereas both objects associated with Swift J1825.7+7215 are type 2 Seyferts.

The case of Swift J1114.3+2020 SW and NE is somewhat different (see Fig. 13). We identify NE as a type 2 LINER galaxy at a redshift of $z=0.027$ (see Fig. 16), and object SW as an elliptical galaxy lying at $z=0.025$ (see Fig. 20). There are Chandra observations of this system that reveal resolved X-ray emission from both sources \citep{Koss16}.
There is a clear tail of stellar emission flowing outwards from object NE, which means these two galaxies are colliding. We speculate that this could be a case of a minor merger: object NE is apparently smaller ($\sim$10 arcseconds wide, or $\sim$6 kiloparsecs) and thus less massive than object SW ($\sim$24 arcseconds wide, 13.4 kpc), which would then dominate the gravitational pull. This could enhance the nuclear activity in source NE \citep{Hernquist95}. Further analysis is still needed to confirm this scenario, particularly regarding the inclination of the two galaxies with respect to the line of sight. Besides, a high resolution spectrum could provide information on the direction of movement and velocity of each galaxy.


\begin{figure}
\includegraphics[scale=0.44]{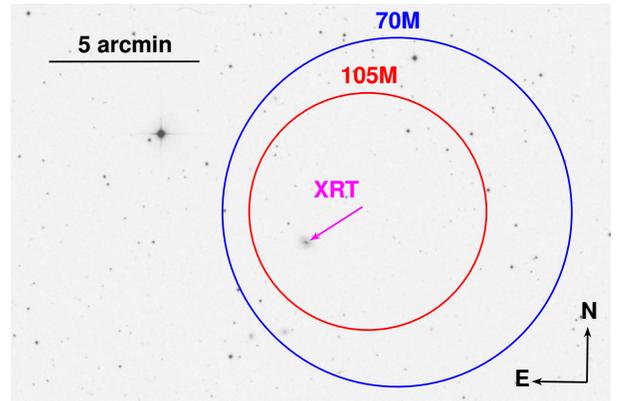}
\caption{Optical DSS-II R-band image of the region corresponding to Swift J0222.3+2509. In blue we show the detection reported in the Swift70 catalogue, with its 90\% confidence error circle. The detection as reported in the Swift105 catalogue, again with its corresponding 90\% confidence error radii, is shown in red. The magenta arrow indicates the soft X-ray counterpart as seen with XRT; a spectrum of the corresponding optical source is reported in this work.}
\end{figure}

\begin{figure}
\includegraphics[scale=0.44]{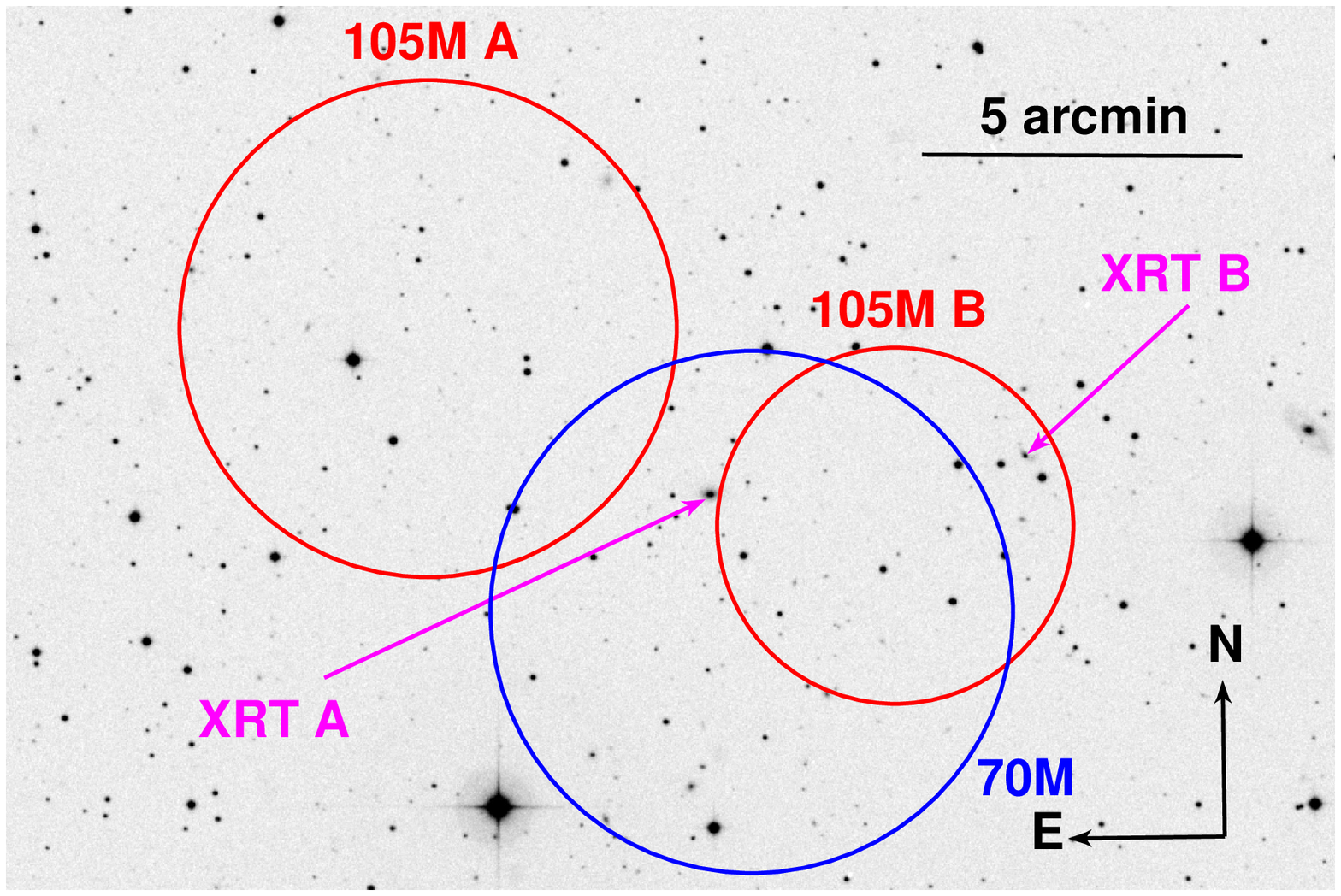}
\caption{Optical DSS-II R-band image of the region corresponding to Swift J0042.9+3016 A and B. In blue we show the detection reported in the Swift70 catalogue, with its 90\% confidence error circle. The detections as reported in the Swift105 catalogue, again with their corresponding 90\% confidence error radii, are shown in red. The magenta arrow indicates the soft X-ray counterparts as seen with XRT; the spectrum of the optical source corresponding to candidate A is reported in this work.}
\end{figure}

\begin{figure}
\includegraphics[scale=0.44]{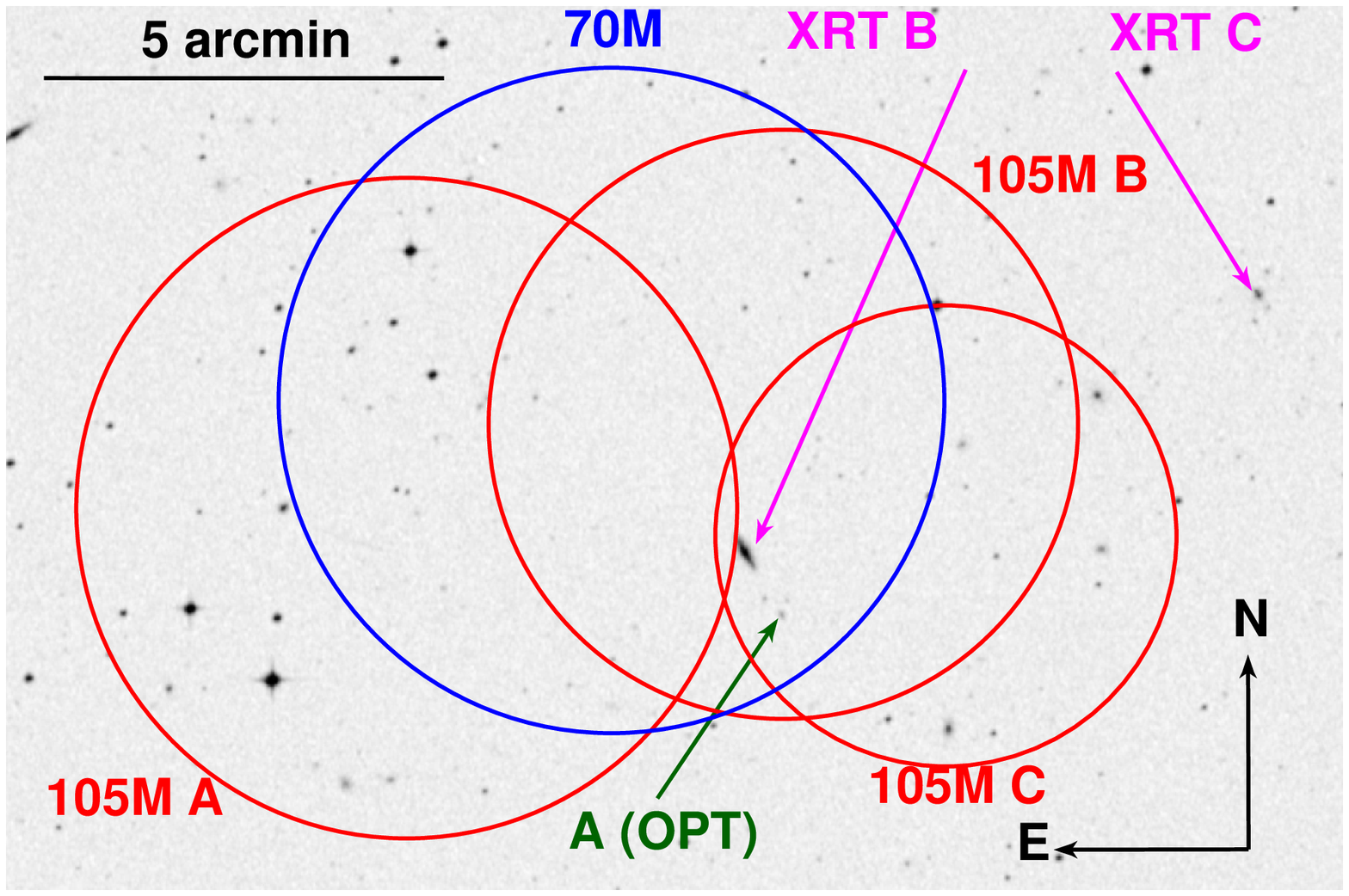}
\caption{Optical DSS-II R-band image of the region corresponding to Swift J0107.7-1137 A, B and C. In blue we show the detection reported in the Swift70 catalogue, with its 90\% confidence error circle. The detections as reported in the Swift105 catalogue, again with their corresponding 90\% confidence error radii, are shown in red. The magenta arrows indicate the soft X-ray counterparts B and C as seen with XRT. Candidate A, although listed in both catalogs, presents no detectable XRT counterpart, and is marked in green. The optical spectra of the three candidates A, B and C are reported in this work.}
\end{figure}

\begin{figure}
\includegraphics[scale=0.44]{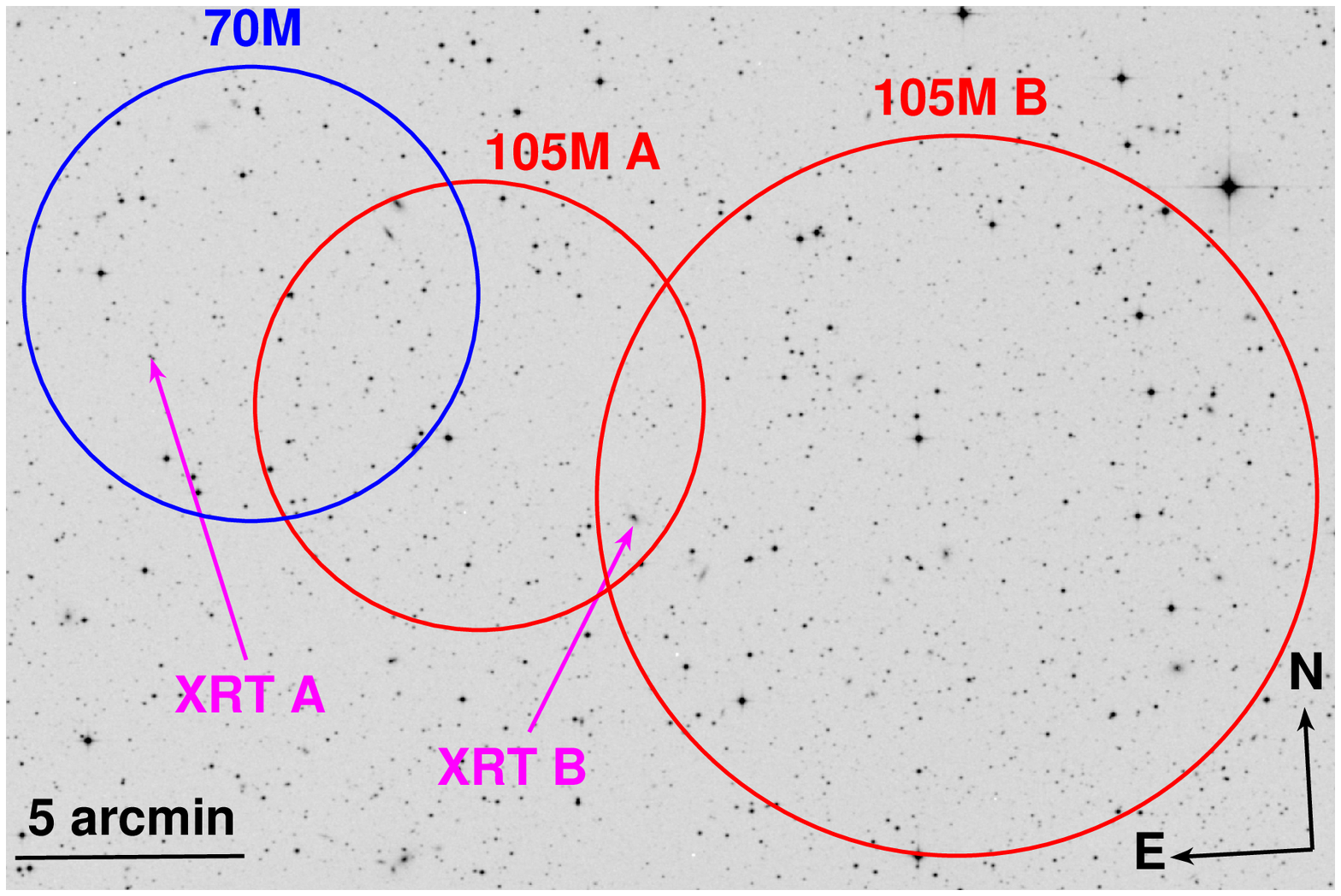}
\caption{Optical DSS-II R-band image of the region corresponding to Swift J0609.5-6245 A and B. In blue we show the detection reported in the Swift70 catalogue, with its 90\% confidence error circle. The detections as reported in the Swift105 catalogue, again with their corresponding 90\% confidence error radii, are shown in red. The magenta arrows indicate the soft X-ray counterparts as seen with XRT; the optical spectrum of candidate B is reported in this work.}
\end{figure}

\begin{figure}
\includegraphics[scale=0.44]{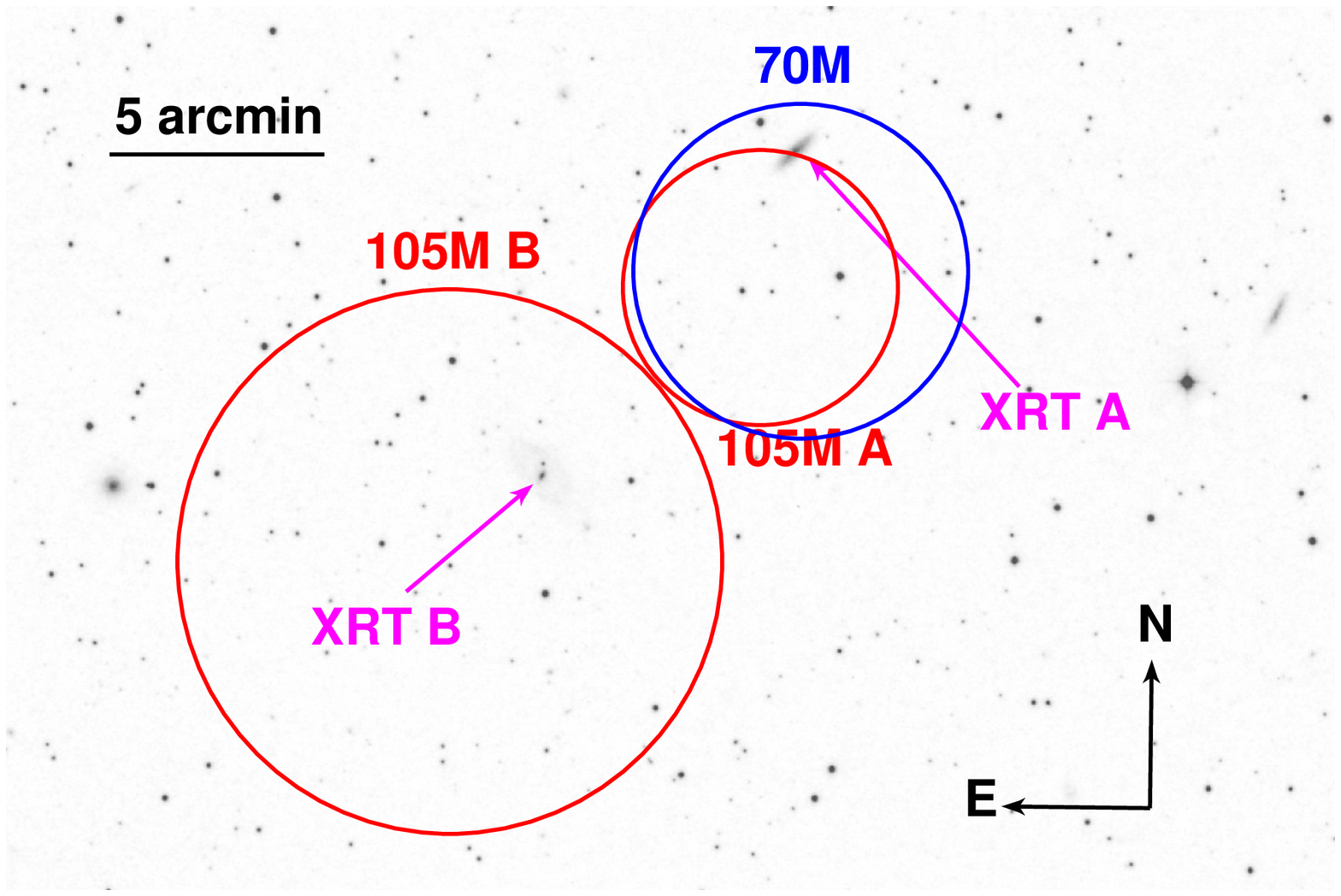}
\caption{Optical DSS-II R-band image of the region corresponding to Swift J1051.2-1704 A and B. In blue we show the detection reported in the Swift70 catalogue, with its 90\% confidence error circle. The detections as reported in the Swift105 catalogue, again with their corresponding 90\% confidence error radii, are shown in red. The magenta arrow indicates the soft X-ray counterparts as seen with XRT; the spectrum of the optical source corresponding to candidate B is reported in this work.}
\end{figure}

\begin{figure}
\includegraphics[scale=0.44]{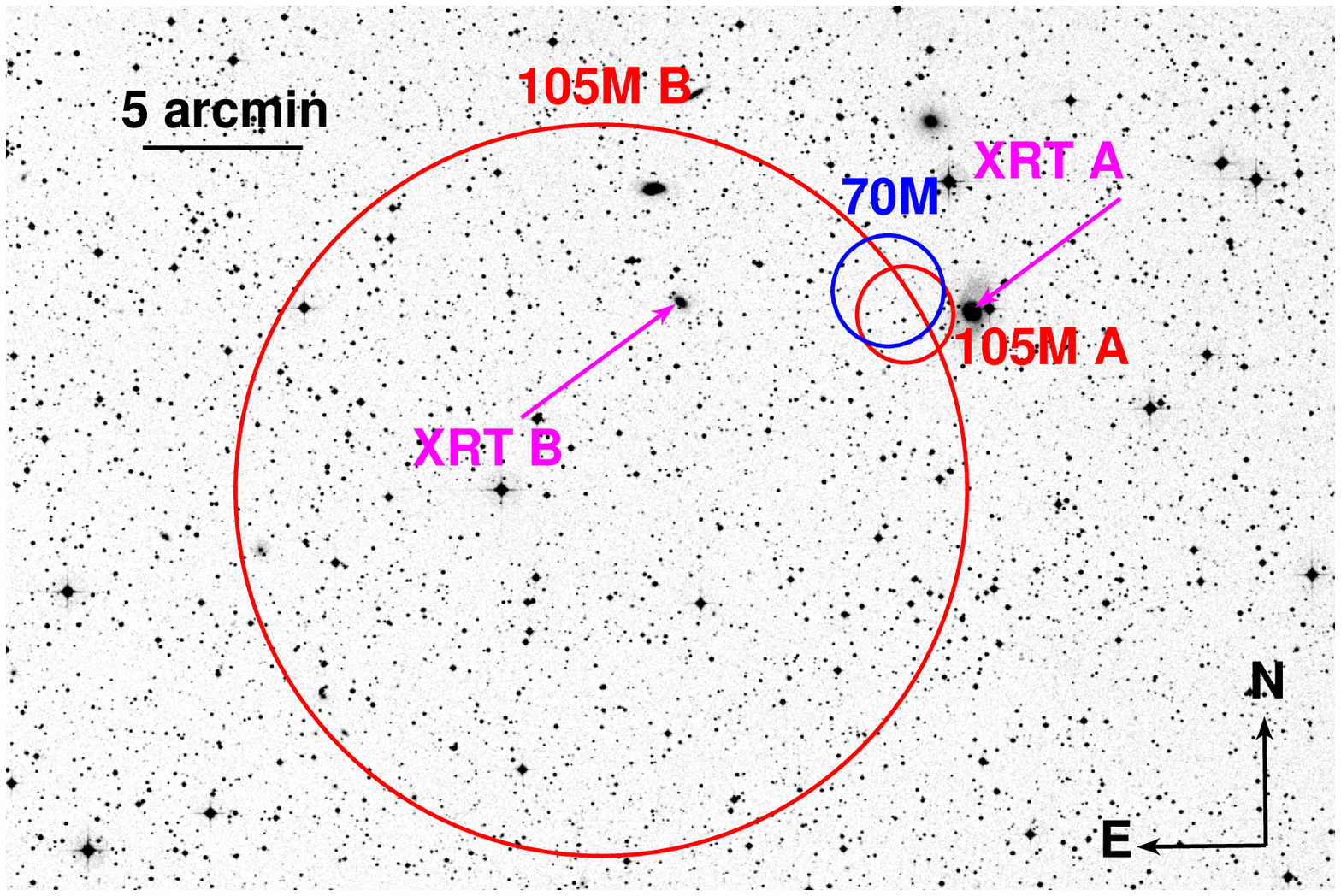}
\caption{Optical DSS-II R-band image of the region corresponding to Swift J1306.5-4025 A and B. In blue we show the detection reported in the Swift70 catalogue, with its 90\% confidence error circle. The detections as reported in the Swift105 catalogue, again with their corresponding 90\% confidence error radii, are shown in red. The magenta arrows indicate the soft X-ray counterparts as seen with XRT; the optical spectrum of candidate B is reported in this work.}
\end{figure}

\begin{figure}
\includegraphics[scale=0.44]{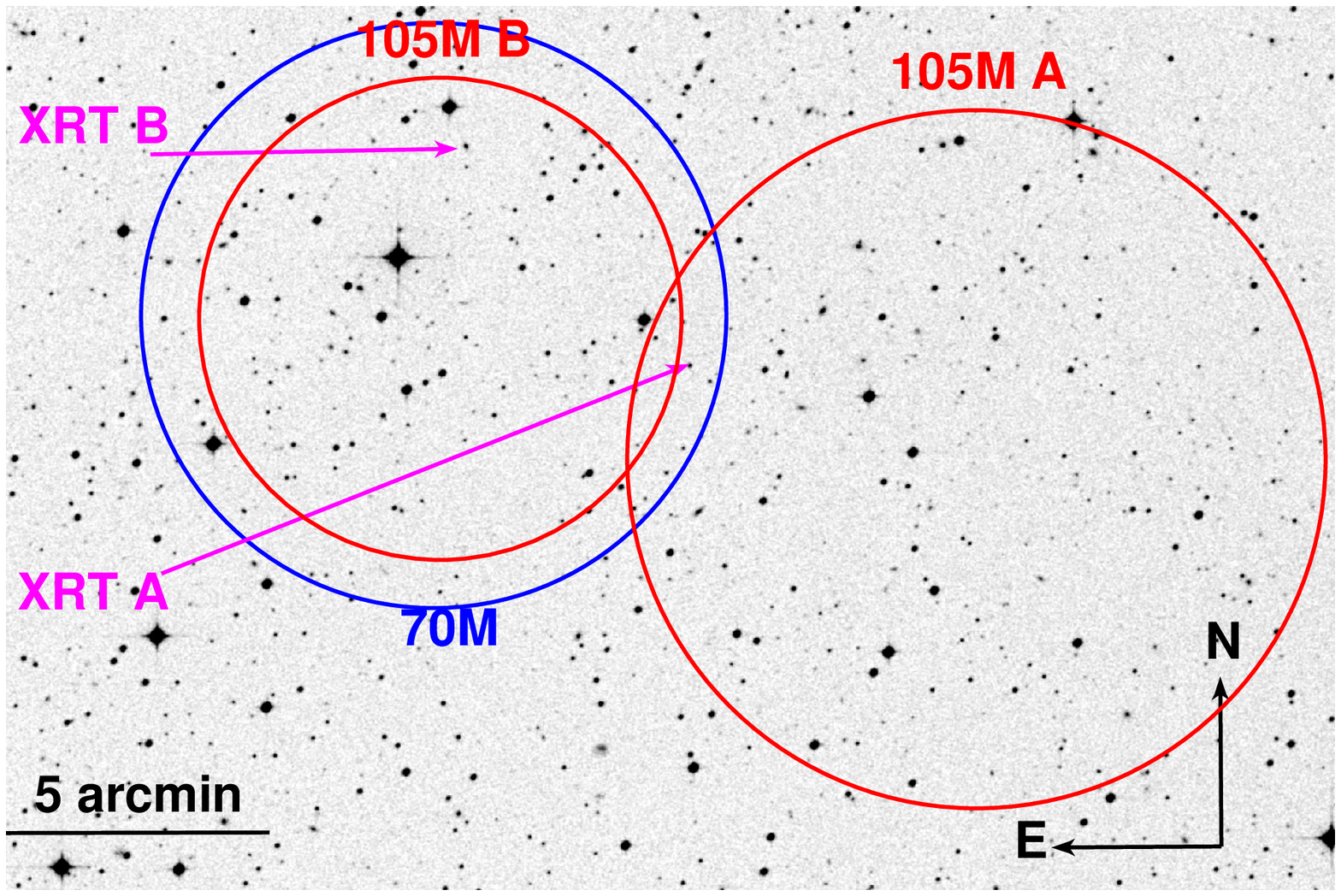}
\caption{Optical DSS-II R-band image of the region corresponding to Swift J1512.2-1053 A and B. In blue we show the detection reported in the Swift70 catalogue, with its 90\% confidence error circle. The detections as reported in the Swift105 catalogue, again with their corresponding 90\% confidence error radii, are shown in red. The magenta arrows indicate the soft X-ray counterparts as seen with XRT; the optical spectrum of candidate B is reported in this work.}
\end{figure}

\begin{figure}
\includegraphics[scale=0.44]{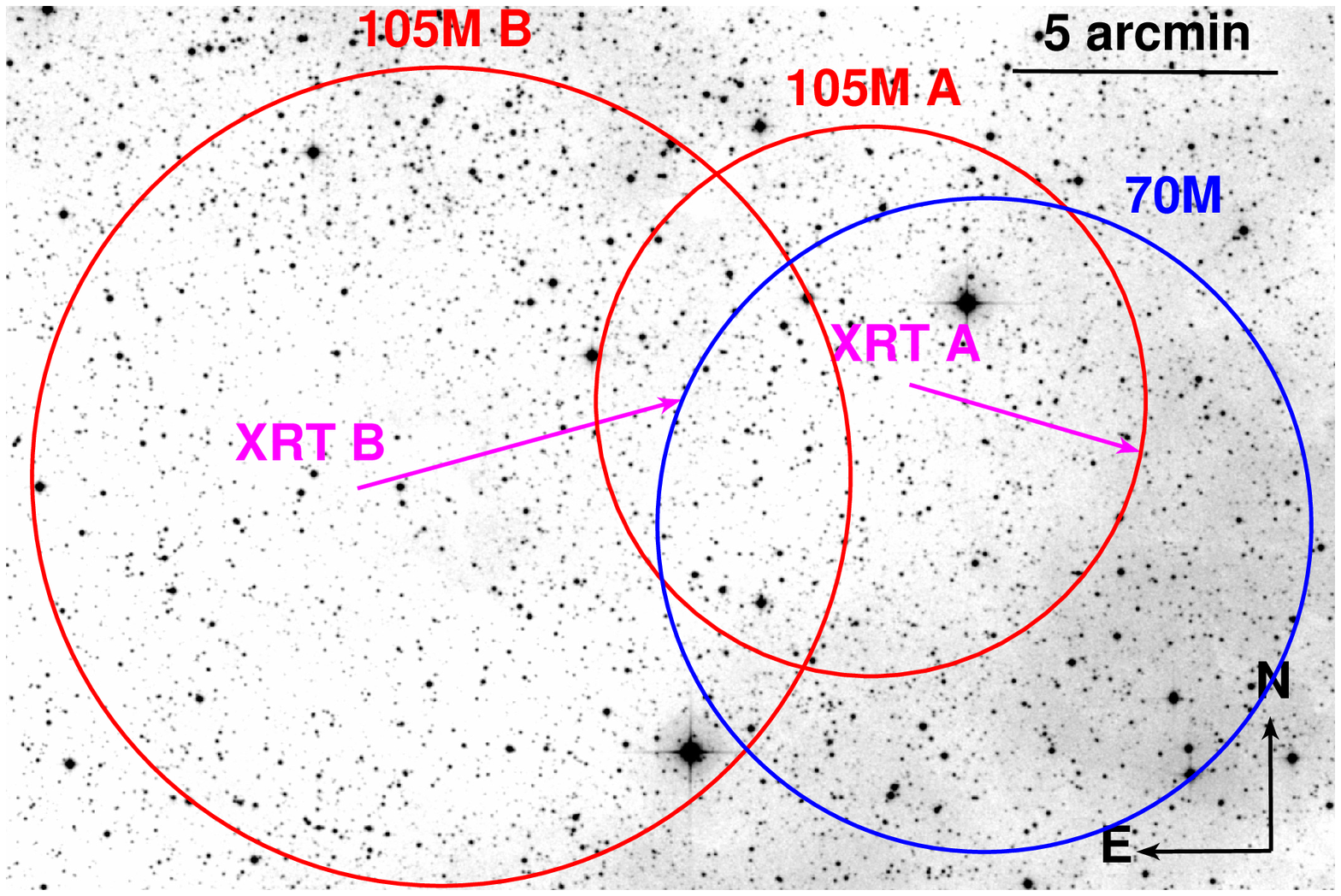}
\caption{Optical DSS-II R-band image of the region corresponding to Swift J2059.6+4301 A and B. In blue we show the detection reported in the Swift70 catalogue, with its 90\% confidence error circle. The detections as reported in the Swift105 catalogue, again with their corresponding 90\% confidence error radii, are shown in red. The magenta arrows indicate the soft X-ray counterparts as seen with XRT; the optical spectra of both candidates are reported in this work.}
\end{figure}

\begin{figure*}[!tbp]
  \centering
    \includegraphics[width=\textwidth]{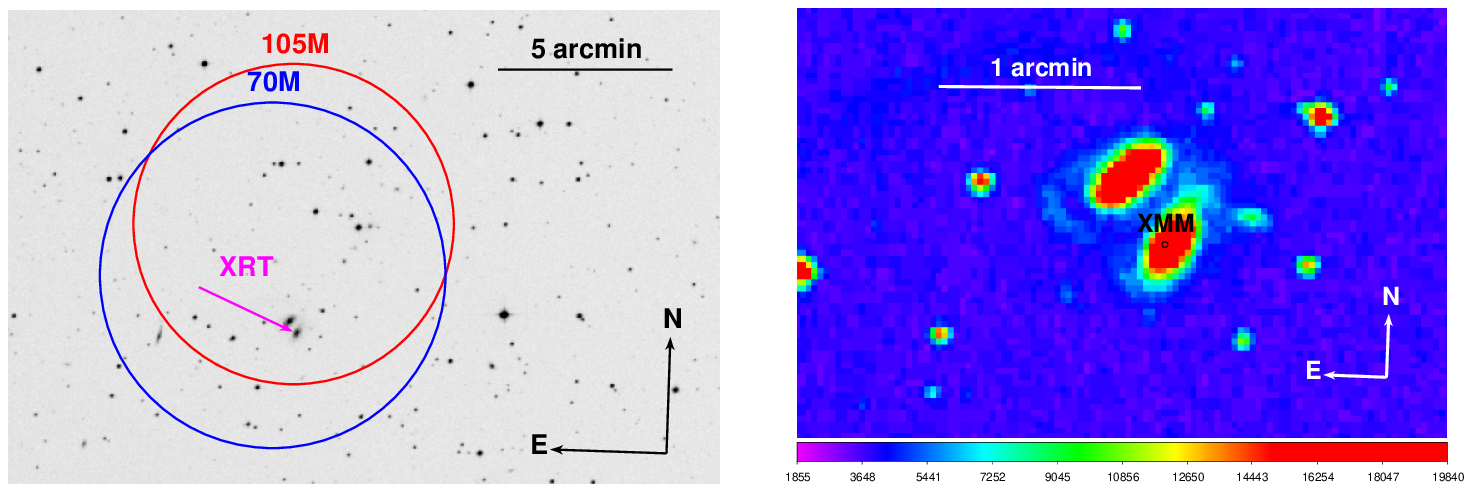}
  \caption{Left panel: Optical DSS-II R-band image of the region corresponding to Swift J0528.1-3933 NE and SW. In blue we show the detection reported in the Swift70 catalogue, with its 90\% confidence error circle. The detection as reported in the Swift105 catalogue, again with its corresponding 90\% confidence error radii, is shown in red. The magenta arrow indicates the soft X-ray counterpart as seen with XMM-Newton. Right panel: Zoomed-in image of the same sky region. The black circle indicates the error radius associated with the soft X-ray counterpart of the source. The optical spectra of both interacting galaxies are reported in this work.}
\end{figure*}

\begin{figure*}[!tbp]
  \centering
    \includegraphics[width=\textwidth]{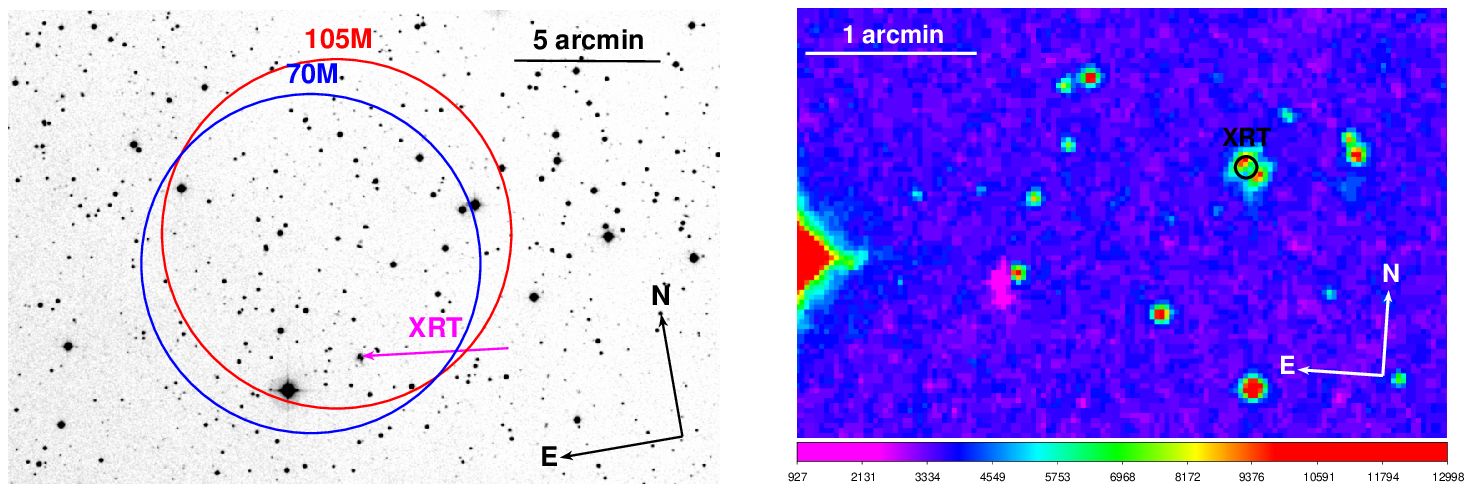}
  \caption{Left panel: Optical DSS-II R-band image of the region corresponding to Swift J1825.7+7215 NE and SW. In blue we show the detection reported in the Swift70 catalogue, with its 90\% confidence error circle. The detection as reported in the Swift105 catalogue, again with its corresponding 90\% confidence error radii, is shown in red. The magenta arrow indicates the soft X-ray counterpart as seen with XRT. Right panel: Zoomed-in image of the same sky region. The black circle indicates the error radius associated with the soft X-ray counterpart of the source. The optical spectra of both interacting galaxies are reported in this work.}
\end{figure*}

\begin{figure*}[!tbp]
  \centering
    \includegraphics[width=\textwidth]{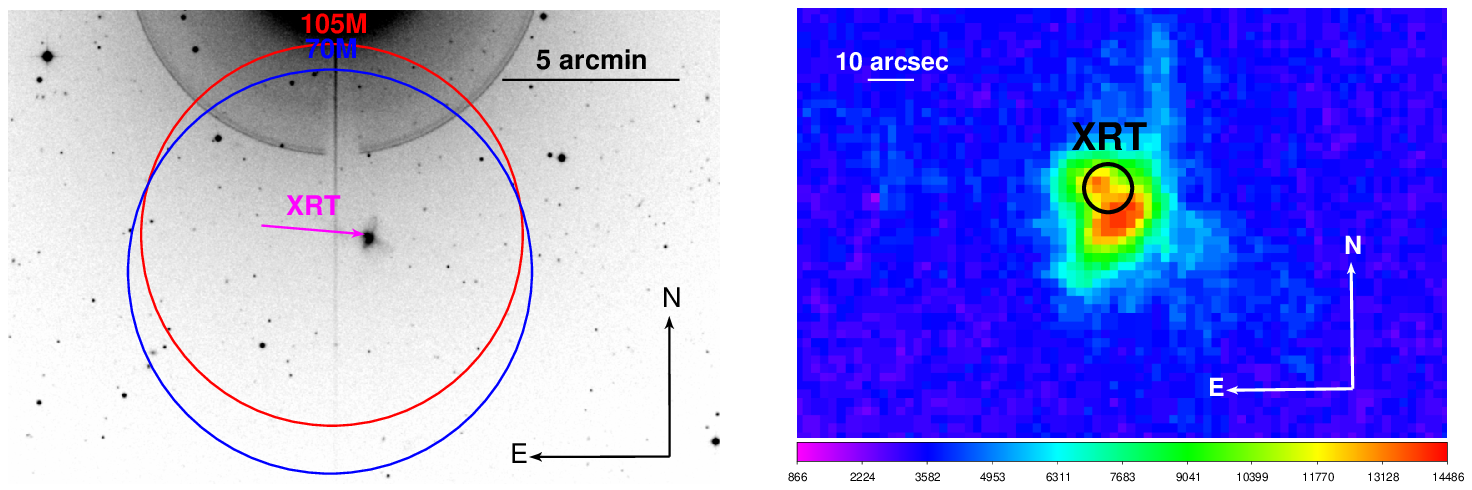}
  \caption{Left panel: Optical DSS-II R-band image of the region corresponding to Swift J1114.3+2020 SW and NE. In blue we show the detection reported in the Swift70 catalogue, with its 90\% confidence error circle. The detection as reported in the Swift105 catalogue, again with its corresponding 90\% confidence error radii, are showin in red. The magenta arrow indicates the soft X-ray counterpart as seen with XRT. Right panel: Zoomed-in image of the same sky region. The black circle indicates the error radius associated with the soft X-ray counterpart of the source. The optical spectra of both interacting galaxies are reported in this work.}
\end{figure*}

\clearpage

\begin{table*}
\caption{Results for all the Galactic sources. In column 1 we give the BAT name, in column 2 its R magnitude from USNO-A2.0, in columns 3 to 7 the identified lines of interest with their respective fluxes and EWs, in column 8 the derived extinction affecting the source (in V magnitudes), in column 9 the distance, and in columns 10 and 11 the X-ray flux and luminosity as observed with BAT. The extinction reported in column 8 was obtained with a Balmer decrement of 2.86 only when the observed H$\alpha$/H$\beta$ ratio was higher than this value. If the observed ratio was lower than 2.86, then we chose to assume that the expected Balmer decrement is 1.0 (see text). Note that all sources are CVs, with the exception of Swift J0428.2-6704A, which is a LMXB. The distance for the latter object was taken from \citet{Strader16}.}
\setlength{\tabcolsep}{2pt}
\small
\begin{tabular}{|c|c|ccccc|c|c|c|c|}
\hline\hline
  \multicolumn{1}{|c|}{Object} &
  \multicolumn{1}{c|}{Mag.} &
  \multicolumn{1}{c}{Line} &
  \multicolumn{1}{c}{Avg. Flux} &
  \multicolumn{1}{c}{Err. Flux} &
  \multicolumn{1}{c}{Avg. EqW.} &
  \multicolumn{1}{c|}{Err. EqW.} &
  \multicolumn{1}{c|}{Av} &
  \multicolumn{1}{c|}{Dist.} &
\multicolumn{1}{c|}{BAT flux}  & 
  \multicolumn{1}{c|}{$\mathrm{L}_X$} \Tstrut  \\
  
  \multicolumn{1}{|c|}{} &
  \multicolumn{1}{c|}{USNO} &
  \multicolumn{1}{c}{} &
  \multicolumn{1}{c}{$\left[\frac{erg\,\times\,10^{-14}}{cm^2\,s}\right]$} &
  \multicolumn{1}{c}{$\left[\frac{erg\,\times\,10^{-14}}{cm^2\,s}\right]$} &
  \multicolumn{1}{c}{[\AA]} &
  \multicolumn{1}{c|}{[\AA]} &
  \multicolumn{1}{c|}{[mag]} &
  \multicolumn{1}{c|}{[pc]} &
  \multicolumn{1}{c|}{$\left[\frac{erg\,\times\,10^{-12}}{cm^2\,s}\right]$} &
  \multicolumn{1}{c|}{$\left[\frac{erg\,\times\,10^{32}}{s}\right]$} \Bstrut \Bstrut  \\

\hline\hline
  2PBC J0819.2-2508 & 19.0 & H$\alpha$ &  0.092 & 0.004 & -71.98 & -6.10 & - & 1000 & 7.0 &  8.4 \Tstrut \\ 
   &  & H$\beta$  & 0.113 & 0.002 & -54.01 & -1.53 & &  &  &   \\ 
   &  & HeII & 0.045 & 0.003 & -17.46 & -1.34 &  &  &  &   \\
   \hline
  Swift J0503.7-2819 & 18.0 & H$\alpha$ &  0.721 & 0.020 & -49.93 & -2.26 & 0.3 & 830 & 8.23 & 6.8\\
   &  & H$\beta$ &  0.662 & 0.017 & -39.35 & -1.62 &  &  &  &   \\
   &  & HeII & 0.410 & 0.093 & -27.46 & -2.34 &  &  &  &   \\
   \hline
  Swift J0535.2+2830 & 18.2 & H$\alpha$ &  0.249 & 0.007 & -19.08 & -0.76 & 0.78 & 730 & 15.07 &  9.6\\
   &  & H$\beta$  & 0.068 & 0.003 & -8.82 & -0.45 &  &  &  &   \\
   &  & HeII & 0.049 & 0.013 & -5.64 & -1.54 &  &  &  &   \\
   \hline
  Swift J2059.6+4301B & 18.4 & H$\alpha$ &  0.263 & 0.009 & -5.35 & -0.23 & 2.5 & 360 & 4.94 & 0.8 \\
   &  & H$\beta$  & 0.116 & 0.009 & -4.34 & -0.35 &  &  &  &   \\
   &  & HeII & 0.165 & 0.032 & -6.29 & -1.03 &  &  &  &  \Bstrut  \\
   \hline
   \hline
   \hline
  Swift J0428.2-6704A & 17.1 & H$\alpha$  & 0.811 & 0.017 & -30.32 & -0.85 & & 2400 & 14.81 & 102.1 \\
   &  & H$\beta$ &  0.114 & 0.010 & -19.75 & -2.00 &  &  &  &   \\
\hline
   \hline

\hline\end{tabular}

\end{table*}

\begin{figure*}
\centering
\includegraphics[scale=0.94]{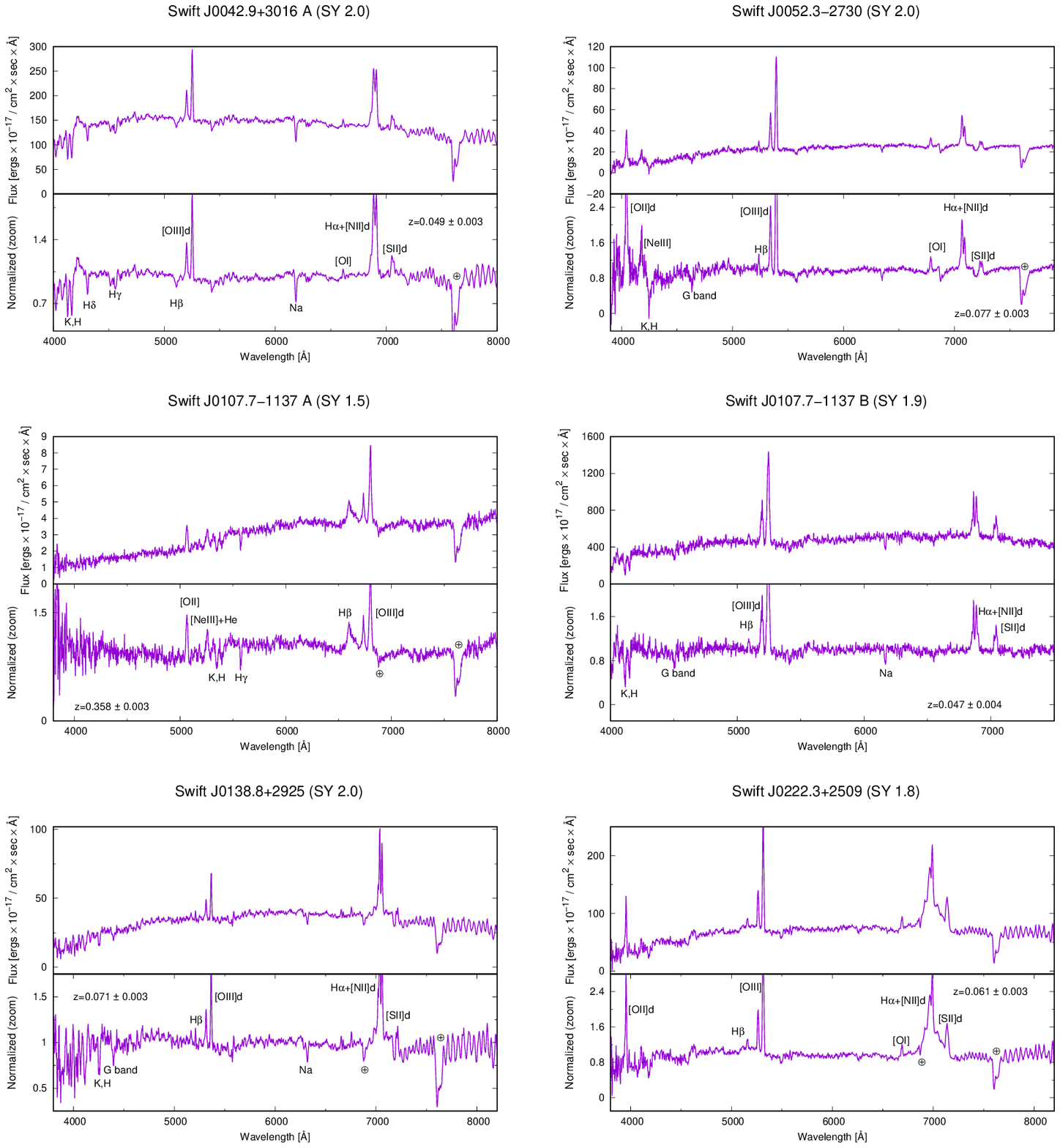}
\caption{Optical spectra obtained for the sample of emission line galaxies presented in this paper. Upper panels show the spectra corrected for Galactic absorption, while lower panels show the spectra with normalised flux. The normalisation was performed through fitting a second order spline cubic function to the continuum and subtracting it. Absorption lines or bands present at 6869 \AA, 7186 \AA, and 7605 \AA\ are telluric and are labeled with an $\oplus$. Any doublets, if present, are marked with a d.}
\end{figure*}

\begin{figure*}
\centering
\includegraphics[scale=0.94]{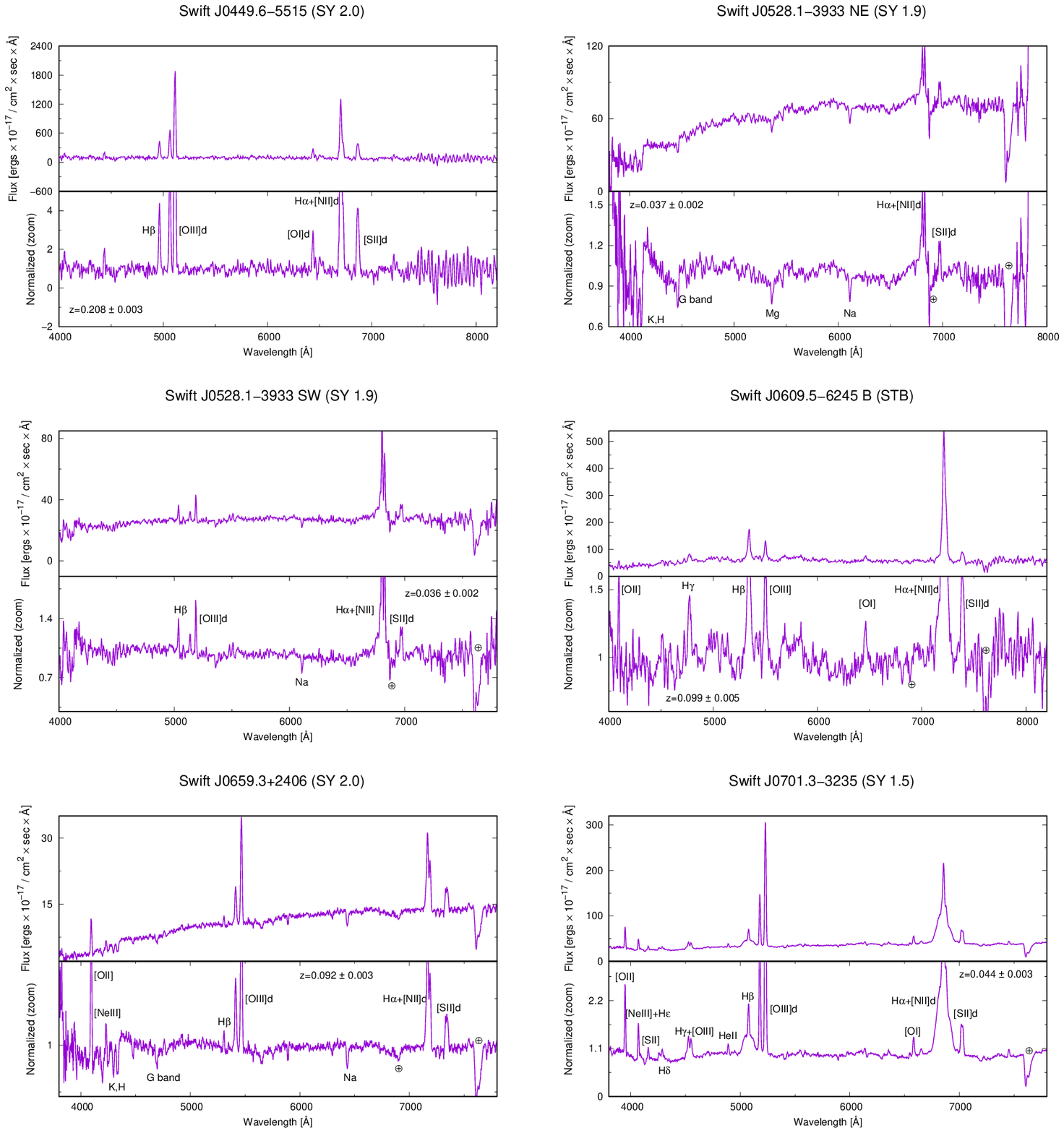}
\caption{Optical spectra obtained for the sample of emission line galaxies presented in this paper. Upper panels show the spectra corrected for Galactic absorption, while lower panels show the spectra with normalised flux. The normalisation was performed through fitting a second order spline cubic function to the continuum and subtracting it. Absorption lines or bands present at 6869 \AA, 7186 \AA, and 7605 \AA\ are telluric and are labeled with an $\oplus$. Any doublets, if present, are marked with a d.}
\end{figure*}

\begin{figure*}
\centering
\includegraphics[scale=0.94]{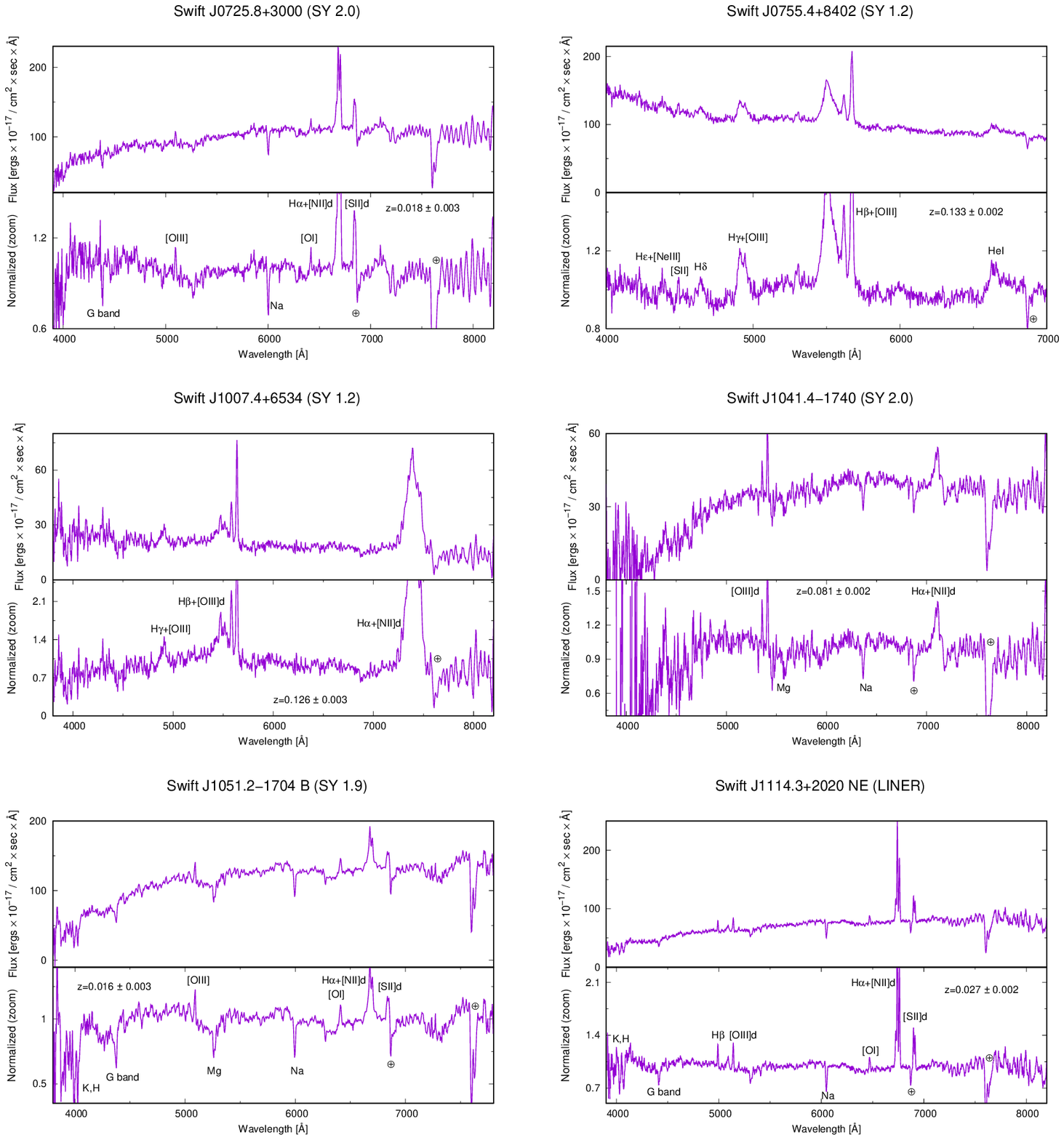}
\caption{Optical spectra obtained for the sample of emission line galaxies presented in this paper. Upper panels show the spectra corrected for Galactic absorption, while lower panels show the spectra with normalised flux. The normalisation was performed through fitting a second order spline cubic function to the continuum and subtracting it. Absorption lines or bands present at 6869 \AA, 7186 \AA, and 7605 \AA\ are telluric and are labeled with an $\oplus$. Any doublets, if present, are marked with a d.}
\end{figure*}

\begin{figure*}
\centering
\includegraphics[scale=0.94]{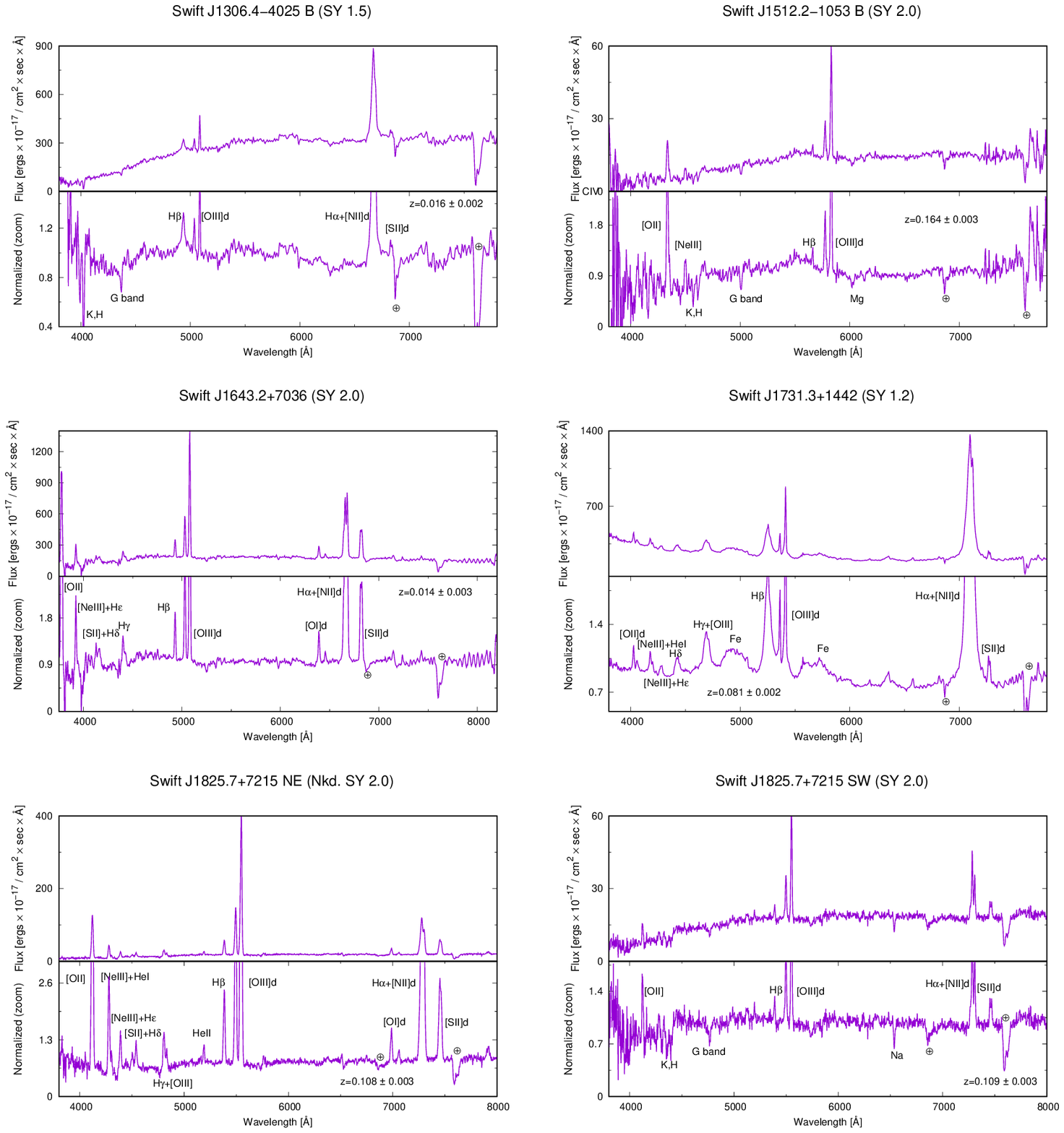}
\caption{Optical spectra obtained for the sample of emission line galaxies presented in this paper. Upper panels show the spectra corrected for Galactic absorption, while lower panels show the spectra with normalised flux. The normalisation was performed through fitting a second order spline cubic function to the continuum and subtracting it. Absorption lines or bands present at 6869 \AA, 7186 \AA, and 7605 \AA\ are telluric and are labeled with an $\oplus$. Any doublets, if present, are marked with a d.}
\end{figure*}

\begin{figure*}
\centering
\includegraphics[scale=0.94,angle=90]{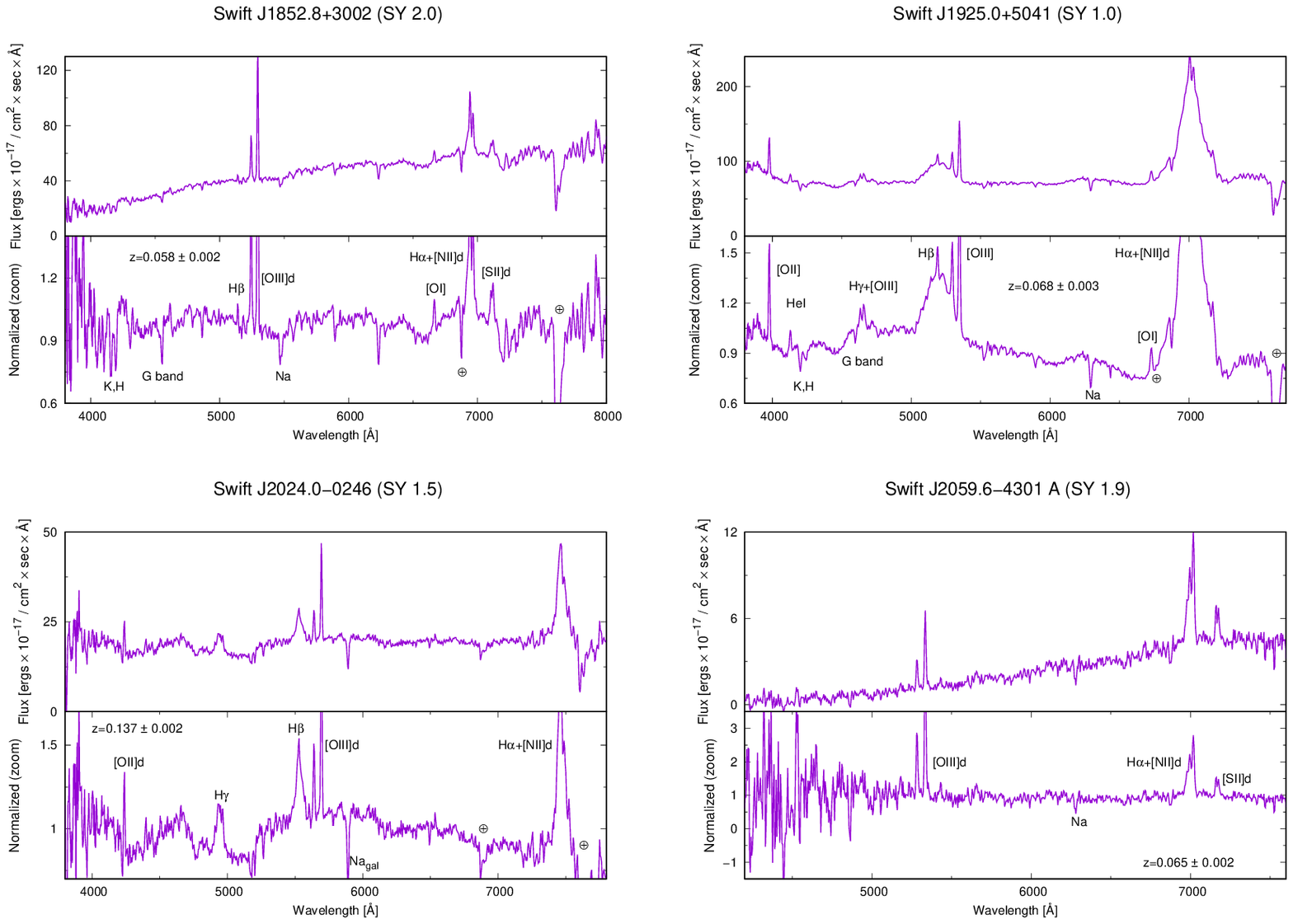}
\caption{Optical spectra obtained for the sample of emission line galaxies presented in this paper. Upper panels show the spectra corrected for Galactic absorption, while lower panels show the spectra with normalised flux. The normalisation was performed through fitting a second order spline cubic function to the continuum and subtracting it. Absorption lines or bands present at 6869 \AA, 7186 \AA, and 7605 \AA\ are telluric and are labeled with an $\oplus$. Any doublets, if present, are marked with a d.}
\end{figure*}

\begin{figure*}
\centering
\includegraphics[scale=0.94,angle=90]{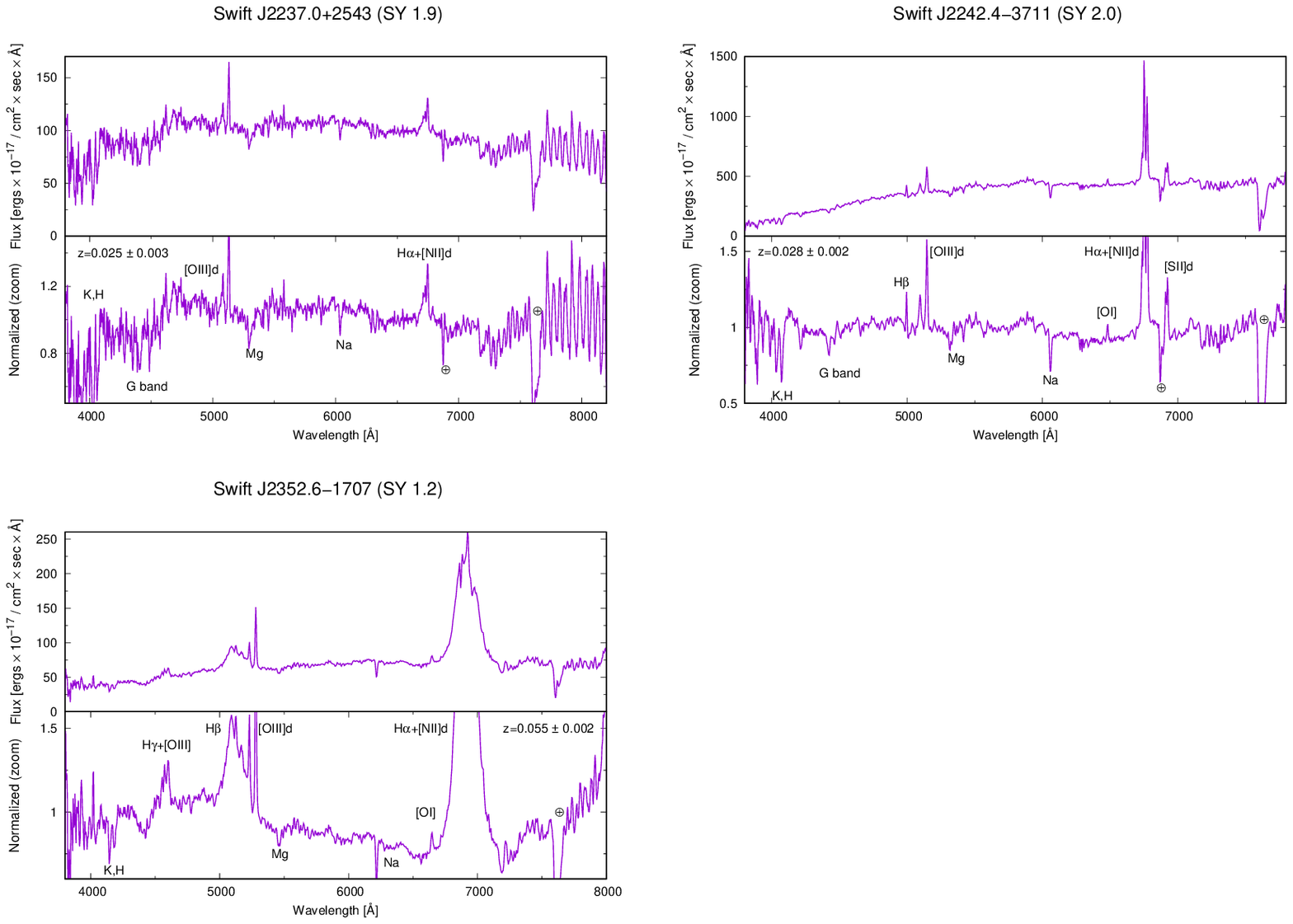}
\caption{Optical spectra obtained for the sample of emission line galaxies presented in this paper. Upper panels show the spectra corrected for Galactic absorption, while lower panels show the spectra with normalised flux. The normalisation was performed through fitting a second order spline cubic function to the continuum and subtracting it. Absorption lines or bands present at 6869 \AA, 7186 \AA, and 7605 \AA\ are telluric and are labeled with an $\oplus$. Any doublets, if present, are marked with a d.}
\end{figure*}

\begin{figure*}
\centering
\includegraphics[scale=0.94]{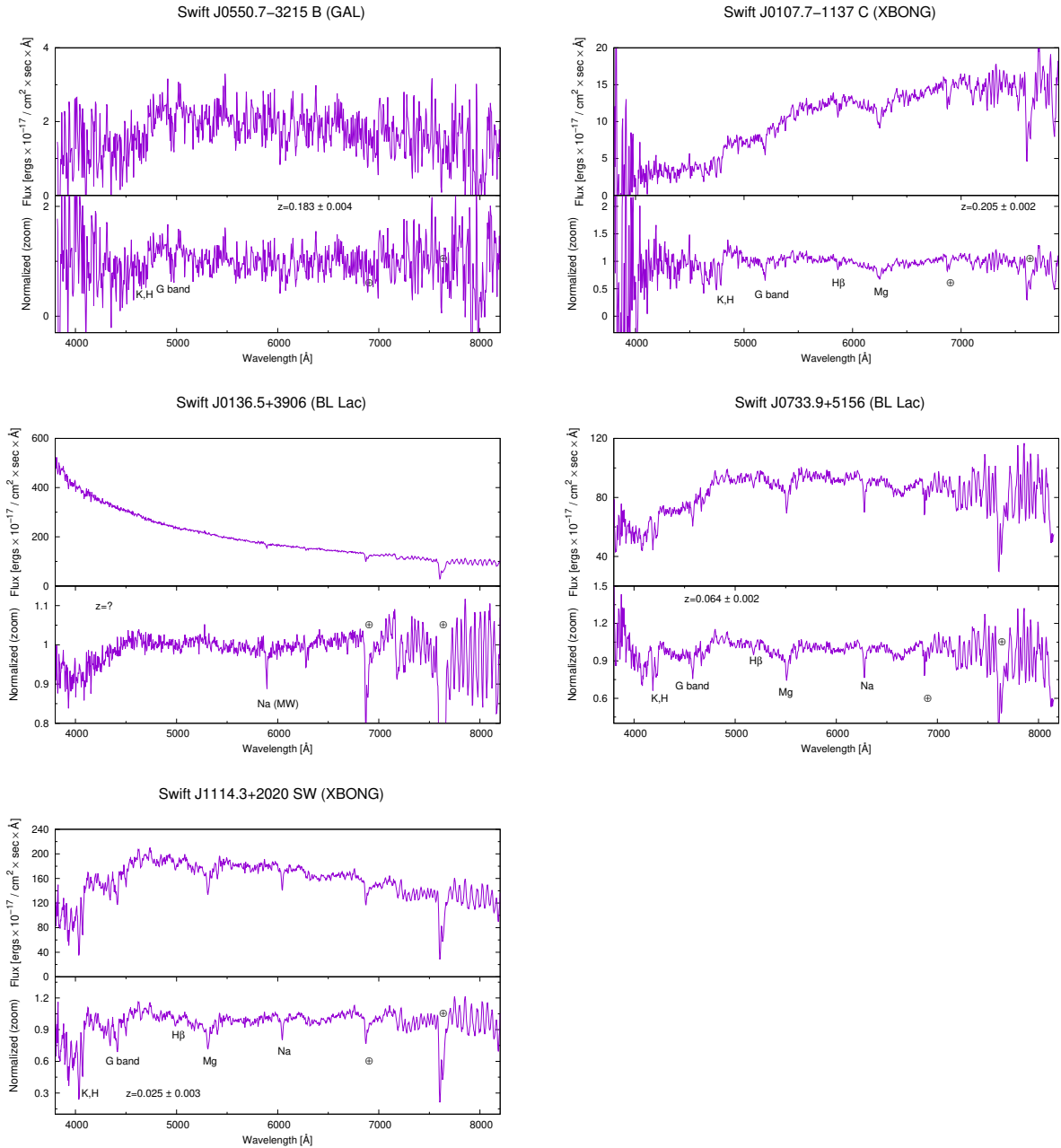}
\caption{Optical spectra obtained for the sample of elliptical galaxies and blazars of BL Lac type presented in this paper. Upper panels show the spectra corrected for Galactic absorption, while lower panels show the spectra with normalised flux. The normalisation was performed through fitting a second order spline cubic function to the continuum and subtracting it. Absorption lines or bands present at 6869 \AA, 7186 \AA, and 7605 \AA\ are telluric and are labeled with an $\oplus$. Any doublets, if present, are marked with a d.}
\end{figure*}

\begin{figure*}
\centering
\includegraphics[scale=0.94]{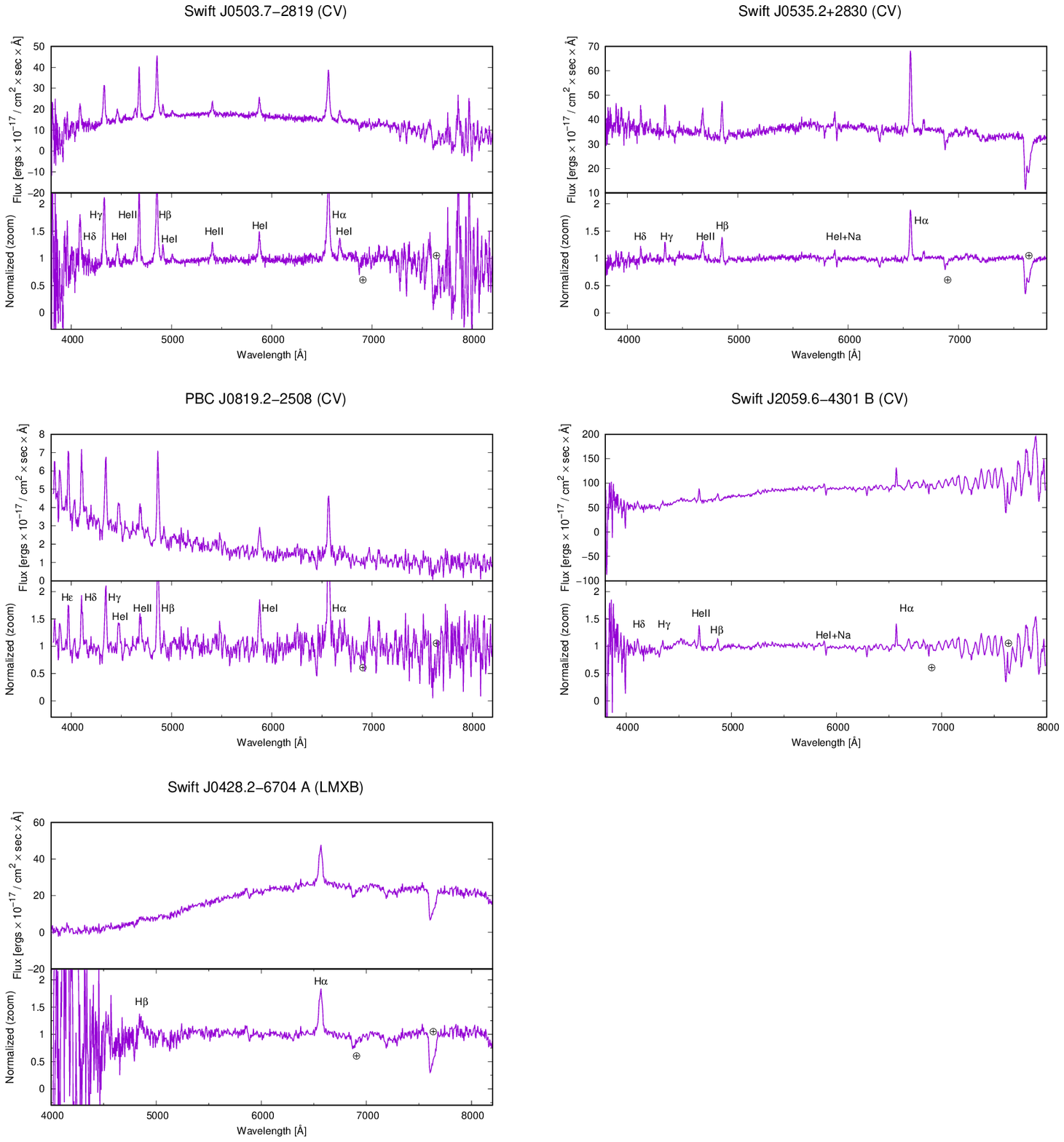}

\caption{Optical spectra obtained for the sample of galactic sources presented in this paper. Upper panels show the observed spectra, while lower panels show the spectra with normalised flux. The normalisation was performed through fitting a second order spline cubic function to the continuum and subtracting it. Absorption lines or bands present at 6869 \AA, 7186 \AA, and 7605 \AA\ are telluric and are labeled with an $\oplus$. Any doublets, if present, are marked with a d.}
\end{figure*}

\section{Conclusions}

In this work, we identified and characterised the proposed optical counterparts of 35 BAT hard X-ray sources that were previously unidentified or poorly studied; all of them are present in the Swift70 catalogue, with the only exception of 2PBC J0819.2-2508. We used optical data taken at 5 different telescopes with apertures ranging from 1.52\,m to 3.58\,m, together with one spectrum from archival data of a spectroscopic survey (performed with a telescope with an aperture of 1.2\,m). Here we summarise our main results.

   \begin{enumerate}
   
         \item We identified the 41 optical counterparts proposed for the aforementioned sample of 35 BAT detections: this allowed us to pinpoint their main spectroscopic characteristics and to associate them (or disprove any association) with the corresponding high-energy emission.
         
      \item The majority of the sample (36 sources) is made of extragalactic objects. In particular, we found 29 Seyfert galaxies (16 of type 1, and 13 of type 2), three elliptical galaxies, two blazars of the BL Lac type, one LINER and one starburst galaxy. We also determined the redshifts for 23 objects of this sample. This further supports and strengthens our statement that optical spectroscopy is a strong tool for the identification of hard X-ray AGNs.
            
      \item Three BAT detections correspond to pairs of interacting galaxies, thus supporting the scenario in which a merger event can trigger nuclear activity as seen in Seyferts. Two of these systems, Swift J0528.1-3933 and Swift J1825.7+7215, are major mergers involving two obscured (types 1.9 and 2) Seyfert galaxies. The remaining one, Swift J1114.3+2020, consists of two merging galaxies (an elliptical one and a LINER). We speculate this must be a case of a minor merger, since the elliptical galaxy is smaller in size than the LINER.
      
      \item Five objects in our sample are galactic sources: four of them are cataclysmic variables, out of which three possibly host a magnetic white dwarf. The remaining object is a confirmed $\gamma$-ray emitting LMXB hosting an accreting neutron star (Swift J0428.2-6704 A).
      
      

   \end{enumerate}

\begin{acknowledgements}
We thank Silvia Galleti for Service Mode observations at the Loiano 
telescope, and Roberto Gualandi, Ivan Bruni and Antonio De Blasi for night 
assistance; we also thank Walter Boschin for coordinating our Service Mode 
observations at the TNG, and Janet Torrealba, H\'ector Ot\'{\i}-Floranes and
Felipe Ram\'on-Fox for observations at SPM.
This research has made use of the ASI Space Science Data Center Multimission 
Archive; it also used the NASA Astrophysics Data System Abstract Service, 
the NASA/IPAC Extragalactic Database (NED), and the NASA/IPAC Infrared 
Science Archive, which are operated by the Jet Propulsion Laboratory, 
California Institute of Technology, under contract with the National 
Aeronautics and Space Administration.
This publication made use of data products from the Two Micron All 
Sky Survey (2MASS), which is a joint project of the University of 
Massachusetts and the Infrared Processing and Analysis Center/California 
Institute of Technology, funded by the National Aeronautics and Space 
Administration and the National Science Foundation.
This research has also made use of data extracted from the Six-degree 
Field Galaxy Survey archive, of the SIMBAD and VIZIER databases operated 
at CDS, Strasbourg, France, and of the HyperLeda catalogue operated at 
the Observatoire de Lyon, France.
V.C. and V.M.P.A. acknowledge support from CONACyT research grant No. 280789. 
The data used in this paper were partially based upon observations carried out 
at the Observatorio Astron\'omico Nacional on the Sierra San Pedro M\'artir 
(OAN-SPM) and Observatorio Astrof\'isico Guillermo Haro (OAGH) in Sonora, 
Mexico.

\end{acknowledgements}

%
%

\bibliography{MarchesiniEJ_bibliography} 
\bibliographystyle{spr-mp-nameyear-cnd}


\end{document}